\documentclass[aps,prd,twocolumn,showpacs,preprintnumbers,nofootinbib,amsmath,amssymb]{revtex4}

\usepackage{graphicx}
\usepackage{bm}
\usepackage{amsthm,url}

\newcommand{\R}{\mathbb{R}}
\newcommand{\Z}{\mathbb{Z}}
\renewcommand{\P}{\mathbb{P}}

\renewcommand{\vec}[1]{\mathbf{#1}}

\newcommand{\Mo}{M_{\odot}}
\def\be{\begin{equation}}
\def\ee{\end{equation}}
\def\bi{\begin{itemize}}
\def\ei{\end{itemize}}
\def\bea{\begin{eqnarray}}
\def\eea{\end{eqnarray}}

\def\bn{\begin{enumerate}}
\def\en{\end{enumerate}}
\renewcommand{\d}{\partial}

\begin{document}

\preprint{gr-qc/0512137}

\title{Best chirplet chain: near-optimal detection of gravitational wave chirps} 
\author{\'Eric Chassande-Mottin}
\affiliation{CNRS, Observatoire de la C\^ote d'Azur, ARTEMIS\\ 
BP 4229 06304 Nice Cedex 4 FRANCE}
\email{ecm@obs-nice.fr}
\author{Archana Pai}
\affiliation{INFN, Sezione Roma 1 --- P.le Aldo Moro, 2 00185 Roma ITALIA \footnote{Present affiliation: Max Planck Institut f\"{u}r Gravitationsphysik (AEI),
 Am M\"{u}hlenberg 1, D-14476 Potsdam, Germany}}
\email{Archana.Pai@aei.mpg.de}
\date{\today}

\begin{abstract}
  The list of putative sources of gravitational waves possibly
  detected by the ongoing worldwide network of large scale
  interferometers has been continuously growing in the last years.
  For some of them, the detection is made difficult by the lack of a
  complete information about the expected signal. We concentrate on
  the case where the expected GW is a quasi-periodic frequency
  modulated signal i.e., a chirp. In this article, we address the
  question of detecting an \textit{a priori} unknown GW chirp. We
  introduce a general chirp model and claim that it includes all
  physically realistic GW chirps. We produce a finite grid of template
  waveforms which samples the resulting set of possible chirps. If we
  follow the classical approach (used for the detection of
  inspiralling binary chirps, for instance), we would build a bank of
  quadrature matched filters comparing the data to each of the
  templates of this grid. The detection would then be achieved by
  thresholding the output, the maximum giving the individual which
  best fits the data.  In the present case, this exhaustive search is
  not tractable because of the very large number of templates in the
  grid. We show that the exhaustive search can be reformulated (using
  approximations) as a pattern search in the time-frequency plane.
  This motivates an approximate but feasible alternative solution
  which is clearly linked to the optimal one. The time-frequency
  representation and pattern search algorithm are fully determined by
  the reformulation.  This contrasts with the other time-frequency
  based methods presented in the literature for the same problem,
  where these choices are justified by ``ad-hoc'' arguments. In
  particular, the time-frequency representation has to be {\it
    unitary}. Finally, we assess the performance, robustness and
  computational cost of the proposed method with several benchmarks
  using simulated data.
\end{abstract}
                                

\maketitle

\section{Introduction}
\label{sec:intro}
The worldwide network \cite{detectors} of large scale interferometric
gravitational wave (GW) detectors have started to take data. The
network includes the detectors GEO600, LIGO and TAMA. It will be
completed soon by the upcoming Virgo. The overall sensitivity of these
detectors is continuously improving. Interesting upper-limits for the
amplitude of GWs are being set and the first detection is hopefully
not too far.

A large variety of astrophysical sources are expected to emit GWs in
the observational frequency bandwidth of these detectors. From the
data analysis viewpoint, the detection methodology for these sources
depends on the availability of a reliable and complete model of the
GW.

Generally speaking, the oscillations of the GWs are related to the
orbital, rotational bulk motion of the constituents of the emitting
system.  Since the system loses energy by radiation, or because of
some other physical process involved, its orbital period, and
consequently the GW frequency can vary with time.  In such case, the
emitted GW is a frequency modulated signal i.e., a \textit{chirp}. A
detailed knowledge of the dynamics of the system is required to
describe precisely the characteristics of the GW chirps, in particular
the phase evolution. This may not be always possible as described in the
following examples.

The GW emitted by a coalescing binary of compact objects can be
divided into three phases (inspiral, merger and ringing).  Although
the GW can be obtained accurately in the inspiral phase when the
bodies are well separated \cite{Blanchet02} (using post-Newtonian
expansions) and in the ringing phase after they have merged
\cite{kokkotas99} (using perturbative methods), the in-between merger
phase still defeats both the numerical and analytical efforts
\cite{schutz04} for modelling its highly non-linear regime. For large
mass binaries, the merger phase contributes to a dominant fraction of
the signal-to-noise ratio (SNR) \cite{flanagan98}.  In this case, the
search method has to accommodate the significant lack of signal
information.

Kerr black holes accreting matter from a surrounding magnetized torus
are putative sources of the long gamma-ray bursts (GRBs)
\cite{Putten04}. It is claimed that, the black hole spin energy is
radiated away through GWs along with the GRB. The precise shape of the
emitted waveform would need accurate hydro-dynamical numerical
simulations.

A third example is the GW emitted in the form of the quasi-normal
modes \cite{Ferrari:2003nk} by a newly born hot neutron star (during
the cooling phase which follows the core collapse). Here, the
characteristics of the GW depend on the equation of state of the
proto-neutron star and various physical processes (like neutrino
diffusion, thermalization and cooling) which are currently not known
with accuracy.

All these three examples are expected to emit GW as an
\textit{unmodelled chirp}, the phase information being not (perfectly)
known. Its typical duration in the detector bandwidth is of the order
of a few seconds.

While matched filtering is a well-known and efficient detection
technique when a precise waveform model is available, the lack of
waveform information prevents us from using the same approach. It is thus
natural to advocate for \textit{exploratory searches} (based on
partial information or ``good sense'' models) as opposed to
\textit{targeted} ones (relying on a precise model).

Various strategies \cite{Arnaud:2002su,Pradier:2000,
  Eric2003,Vicere:2001ye,Anderson:2000yy,Anderson:1999pr,Sylvestre:2002ih}
have been designed following this viewpoint, for the detection of
transients of short duration (tenth to hundredth of milliseconds) or
GW \textit{burst}. Such transients are typically from supernovae
core collapses. The notion of a varying frequency is not adequate for
such a small number of cycles. It is thus not meaningful to describe
such transients as chirps. Their detection is a different issue than
the one considered here.

Here, we are interested in exploratory search specifically for
unmodelled chirps. In the past, this question has already been
investigated yielding a detection method, the Signal Track Search
\cite{Anderson:1999pr} (STS). The STS relies on the observation that,
in a time-frequency (TF) representation, a chirp appears as a filiform
pattern and this discriminatory signature can be searched for. A
satisfactory implementation of this phenomenological argument calls
for a proper TF representation (TFR) and pattern
search algorithm. The STS results from ``ad-hoc'' choices for the
above mentioned points.

In this paper, we propose a new method for the detection of unmodelled
chirps. It is based on the same general principles (pattern search in
a TFR) as the STS. Its originality resides in the clear link we
establish between the method (i.e., the choices of TFR and pattern
search) and an optimality criterion.

The paper is organized as follows.  We state the detection problem in
Sec. \ref{sec:problem}. We introduce the general chirp model referred
to as \textit{smooth chirp} and we assume that most physically
realistic GW chirps can be described by this model. The phase of a
smooth chirp is an arbitrary continuous and differentiable function
with bounded first and second derivatives. In Sec. \ref{sec:optimal},
we derive the optimal statistic for detecting a given smooth chirp in
noise, which is usually referred to as \textit{quadrature matched
  filtering}.  The idea is then to apply this statistic for any smooth
chirp, and select the maximum which is associated to the individual
that best fits the data. This maximization has to be done numerically.
To do so, the set of smooth chirp being a continuous set, has to be
discretized. In Sec.~\ref{sec:cc}, we show that grids of templates can
be constructed for smooth chirps using chains of small chirps, we call
\textit{chirplet chains} (CC). We further prove that the grid is
\textit{tight} i.e., any smooth chirp can be closely approximated by a
chirplet chain. The maximization of the statistic over the set of
smooth chirp can be reliably replaced by a maximization of the set of
CCs. However, the number of CCs being very large, the computation of
the quadrature matched filter for all CCs is not tractable.  In Sec.
\ref{sec:maxcc}, we propose a feasible (TF based) procedure for
finding the best CC. We show that the quadrature matched filter can be
reformulated approximately as a path integral computed in the TF
representation given by the discrete Wigner-Ville (WV) distribution.
As a result, the maximization of the statistic over the CCs amounts to
obtaining the TF path of largest integral. We demonstrate that this
kind of problem can be solved efficiently with dynamic programming.
We detail our path search algorithm and we evaluate its computational
cost.  Finally, we compare the resulting algorithm with other methods
in Sec.  \ref{sec:test}. Receiver operating characteristics obtained
in several realistic situations demonstrate the superiority of the
proposed approach.

\section{Smooth chirps in Gaussian noise}
\label{sec:problem}
We introduce a general chirp model which we refer to as \textit{smooth chirp},
\begin{equation}
\label{smooth_chirp}
s(t)=A \cos(\phi(t)+\varphi_0) \quad \text{for $t_0\leq t \leq t_0+T$} \,,
\end{equation}
and $s(t)=0$ outside this interval.

A smooth chirp is characterized by the amplitude $A$, the initial
phase $\varphi_0$ and a \textit{smooth} phase evolution $\phi(t)$ (without
loss of generality, we assume $\phi(t)=0$ at the arrival time
$t=t_0$). We define the term \textit{smooth} as follows. A phase
$\phi(t)$ is smooth if this function and its first three derivatives
are continuous and we have
\begin{align}
\label{smooth_def}
\left|\frac{df}{dt}\right|&\leq \dot{F}&\left|\frac{d^2f}{dt^2}\right|&\leq \ddot{F},
\end{align}
for all $t$ and where $f(t)\equiv (2\pi)^{-1} d\phi/dt$ is the
instantaneous frequency. The chirping rate limits $\dot{F}$ and
$\ddot{F}$ are chosen based on the allowed upper bounds obtained from
general astrophysical arguments on the GW source of interest.  The
chirp model thus includes four parameters $\vec{p}\equiv \{A,
\varphi_0, t_0, \phi(\cdot)\}$ which are not known \textit{a priori}
and need to be determined from the data.

Let the signal be correctly sampled at the Nyquist rate $f_s\equiv
1/t_s$ and let assume that we acquire the data $x_k$ by blocks of $N$
samples. The signal is denoted by $s_k \equiv s(k t_s)$ for $k=0,
\ldots, N-1$ with the duration $T=t_s N$.  The noise $n_k$ is assumed
to be additive white and Gaussian with zero mean and unit variance.
Since the noise of GW detectors is colored, this noise model implies
that a whitening procedure has been already applied to the data.
(Therefore, the signal $s_k$ in Eq.~(\ref{detection}) is a
``whitened'' version of the actual GW signal).

In this initial work, we restrict the smooth chirp model to have a
constant envelope, although GW chirps are generally amplitude
modulated. The constant envelope thus limits the descriptive power of
the model. However, we argue that the model is still reasonable for
many cases\footnote{It is important to stress here that the model
  applies to the ``whitened'' chirp. For inspiralling binary chirps
  crossing the entire detector bandwidth, the envelope of ``whitened''
  chirp is flatter than the original GW signal. For the other cases,
  this fact depends on the location of the chirp within the detector
  band.} and that the phase information plays a major role for
detection of chirps.  We leave the problem of detecting amplitude
modulated chirps for subsequent work.

\section{Optimal detection of a smooth chirp}
\label{sec:optimal}
For each block of $N$ data samples, the signal detection problem is to decide
which statistical hypothesis suits best to the data among the
following two~:
\begin{subequations}
\label{detection}
\begin{align}
(H_0) \quad x_k&= n_k \quad \quad \quad \text{noise only} \\
(H_1) \quad x_k&= s_k + n_k \quad \text{signal+noise} 
\end{align}
\end{subequations}

In practice, this requires thresholding a functional of the data,
commonly referred to as \textit{statistic}. If the statistic crosses
the threshold, $H_1$ is chosen as opposed to $H_0$ and {\it vis a versa}.

Due to the presence of random noise, this decision may not be always
the right one. There are two types of errors associated to this: false
alarms (decide $H_1$ while $H_0$ is present) and false dismissals (the
opposite).  The probabilities of occurrence of these two errors fully
quantify the performance of a given statistic. This information can be
used to rank the large number of possible statistics and to identify
the best one. This is the approach followed by the Neyman-Pearson (NP)
criterion \cite{Kay:1998}: the NP--optimal statistic minimizes one
error probability, while keeping the other fixed to a given value.  To
be precise, in the present case, it minimizes the false dismissal
probability for a fixed false alarm probability.

For simple detection problems, it can be shown that
the likelihood ratio (LR) defined by
$\lambda\equiv\P(\{x_k\}|H_1)/\P(\{x_k\}|H_0)$ is NP--optimal
\cite{Kay:1998}. For 
smooth chirp detection problem described in Eq.~(\ref{detection}),
the LR can be easily obtained if we
assume that the chirp parameters $\vec{p}$ are known in advance. 
When the parameters are not known \textit{a priori}
(which is the situation here), the ideal would be to have a statistic
which is NP--optimal for all values of the parameters. This statistic
is usually referred to as \textit{uniformly most powerful}. However,
it is not guaranteed that such statistic always exists, and even if it does,
it is generally difficult to
obtain.

A sensible solution consists in getting some kind of estimates for the
unknown parameters and then use the LR assuming
that the estimated value is the actual value. If we use
\textit{maximum likelihood} (ML) estimators of the
unknown parameters, the resulting statistic is referred to as
\textit{generalized likelihood ratio test} (GLRT) \cite{Kay:1998} (or
maximum likelihood test in the statistical community).

The GLRT can be shown to be uniformly most powerful in certain cases
\cite{Kay:1998}. For our problem, up to our knowledge, this is an open
question. Strictly speaking, it is thus not correct to qualify the
GLRT as ``optimal'' (as is often done in the literature on GW data
analysis). Nevertheless, we continue this misuse of language
since the GLRT has proven to perform reasonably well and no better
alternative appears to be available.

In the following subsections, we give the derivation of the GLRT
statistic.  We proceed with the maximization of likelihood ratio with
respect to the parameters. Following \cite{Owen:1998dk}, we note that
out of the four parameters, $A$, $\varphi_0$ and $t_0$ are
\textit{extrinsic} parameters (known as kinematical or dynamical
parameters) whereas $\phi(\cdot)$ is an \textit{intrinsic} parameter
(which determines the shape of the chirp waveform). On the basis of
this distinction, the maximization over the extrinsic parameters can
be treated in a simple manner whereas the computation of the ML
estimate of the intrinsic parameter requires a more sophisticated
numerical treatment.

\subsection{Maximize the likelihood ratio: $A$ and $\varphi_0$ }
In this subsection, we maximize the LR with respect to $A$ and $\varphi_0$.
In case of Gaussian noise, it is more convenient to use log-likelihood ratio
(LLR) which is expressed by
\begin{equation}
\label{llr}
\Lambda(x;\vec{p})\equiv \ln \lambda = \sum_{k=0}^{N-1}x_k s_k-\frac{1}{2}\sum_{k=0}^{N-1}s_k s_k.
\end{equation}

We introduce $\bar{s}_k \equiv \cos(\phi_k+\varphi_0)$
(such that $s_k=A\bar{s}_k$) with the norm ${\cal N}\equiv
\sum_{k=0}^{N-1}\bar{s}^2_k$.

The maximization of the LLR $\Lambda(x;\vec{p})$ over $A$ is
straightforward and gives the expression of the ML estimate of the
amplitude, namely $\hat{A}=\sum_{k=0}^{N-1}x_k \bar{s}_k/{\cal N}$.
Inserting this expression into Eq.~(\ref{llr}), we obtain
\begin{equation}
  \Lambda(x;\{\hat{A},\varphi_0,t_0,\phi(\cdot)\})=\frac{1}{2{\cal N}}\left(\sum_{k=0}^{N-1} x_k \bar{s}_k\right)^2.
\label{Eq:lambda}
\end{equation}

The analytical maximization of the LLR over $\varphi_0$ deserves a
little more attention. The same calculation has been performed for the
detection of chirps from inspiralling binaries
\cite{Sathyaprakash:1991mt,Owen:1995tm} but it is based on the
assumption that $\cal N$ is independent of $\varphi_0$ which is not
valid in the context of arbitrary chirps. In Appendix \ref{sec:llr},
we detail this calculation and discuss the validity of this
assumption.

We express the resulting statistic
$\ell(x;t_0,\phi)\equiv\Lambda(x;\{\hat{A},\hat{\varphi}_0,t_0,\phi(\cdot)\})$
using the following notations for the cross-correlation of the data
with the two quadrature waveforms,
\begin{align}
\label{cos_sin}
 x_c &\equiv \sum_{k=0}^{N-1}x_k \cos\phi_k  
&x_s &\equiv \sum_{k=0}^{N-1}x_k \sin\phi_k \,,
\end{align}
and for the norms and cross-products of $\cos\phi_k$ and $\sin\phi_k$,
\begin{subequations}
\label{normal_var}
\begin{gather}
n_c \equiv \sum_{k=0}^{N-1}\cos^2\phi_k \quad\quad
n_s \equiv \sum_{k=0}^{N-1}\sin^2\phi_k\\
n_x \equiv \sum_{k=0}^{N-1}\cos\phi_k \sin\phi_k \,. 
\end{gather}
\end{subequations}

We distinguish two cases. In the degenerate case where the two
quadrature waveforms are linearly dependent ($\phi_k$ is a constant),
${\cal O}\equiv n_c n_s-n_x^2$ vanishes and we have
\begin{equation}
\label{eq:stat2}
\ell(x;t_0,\phi)=(x_c^2+x_s^2)/(2N).
\end{equation}

Otherwise ${\cal O}>0$, the optimal statistic is
\begin{equation}
\label{eq:stat1}
\ell(x;t_0,\phi)=(n_s x_c^2 - 2 n_x x_c x_s + n_c x_s^2)/(2{\cal O}),
\end{equation}
and is commonly referred to as \textit{quadrature matched filtering},
(see Appendix \ref{sec:llr}).

\subsection{Maximize the likelihood ratio: $\phi$ and $t_0$}
The statistic $\ell$ results from a quadratic combining of the
cross-correlations defined in Eq.~(\ref{cos_sin}). It can be seen as a
``generalized dot-product'' and can be related to a ``distance''
measuring the discrepancy between the data and \textit{template}
waveforms (or, in short, templates) defined by the phase $\phi$ (see
Eq.~(\ref{statistic_GS2})).  Maximizing $\ell$ over $\phi$ is
equivalent to minimizing this distance.

The expression in Eq.~(\ref{eq:stat1}) is for a given known phase $\phi$.
If the phase is unknown but belongs to the set of smooth chirps, then
we need to minimize the distance within this feasible set. In other
words, we need to find that smooth chirp which best fits the data
i.e., find
\begin{equation}
\label{max_llr}
\ell_{\mathrm{max}}(x;t_0)=\max_{\text{all smooth chirps}}\{\ell(x;t_0,\phi)\}. 
\end{equation}

This maximization is difficult to tackle analytically and has to be
done numerically. The set of smooth chirps is a continuous set and
hence not easy to manipulate numerically without discretizing it. For
this purpose, we introduce \textit{chirplet chains}, which we discuss
in the next section.

As described earlier, we process the data stream block-wise. We
compute the statistic independently for each block. The maximization
over $t_0$ is obtained by comparing $\ell_{\mathrm{max}}$ for
neighbouring blocks and selecting the maximum. The ML estimate of
$t_0$ is then given by the starting time of the corresponding block.
The period separating two successive starting times thus defines the
resolution of the estimate. If required, this resolution can be
improved by increasing the overlap between two neighbouring blocks.

We now concentrate on the maximization of $\ell(x;\phi)$
over $\phi$ in a given block. In the following, we remove $t_0$ from
the arguments of $\ell$ to keep the notations simple.

\section{Chirplet chains: a tight template grid for smooth chirps}
\label{sec:cc}
In this section, we show that chirplet chains (CCs) can be used to
construct template grids for smooth chirps. CCs are based on the
simple geometrical observation: broken lines give good approximations
of smooth curves. CCs are signals whose (instantaneous) frequency is a
broken line. We verify that they are good approximation of the
frequency curve of an arbitrary smooth chirp. We obtain the conditions
ensuring that, for any smooth chirp, there always exists a
sufficiently close CC. If these conditions are satisfied, the set of
the CCs forms a tight template grid which can be used to search for an
unknown smooth chirp.  Finally, we examine the implementation of such
grid for the toy (but realistic) model given by the inspiralling
binary chirp.

\subsection{Chirplet chains: piecewise linear frequency}
\label{sec:cc_def}
\begin{figure}
  \centerline{\includegraphics[width=\columnwidth]{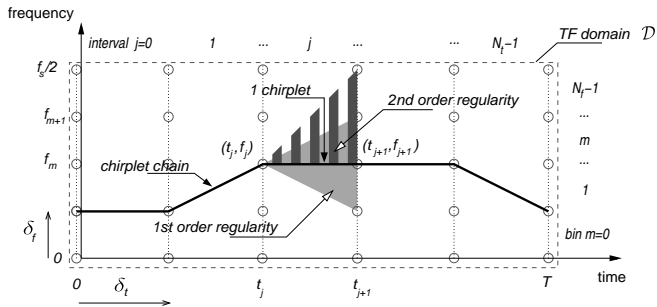}}
  \caption{\label{fig:cc} \textbf{Chirplet chains ---} The TF domain
    of interest ${\cal D}$ is tiled by a regular TF grid of $N_t$ time
    intervals and $N_f$ frequency bins. A chirplet is a short piece of
    signal whose frequency varies linearly between two successive
    nodes of the grid. It is thus represented by a line joining the
    grid nodes. The slope of the chirplet frequency is limited
    (triangular region in light gray, here $N_r'=1$).  We chain the
    chirplets, imposing the continuity of the chain and limiting the
    difference between the slopes of two consecutive chirplets
    (triangular region with dark gray stripes, here $N_r''=1$).
    Admissible chirplets in time interval $j$ belong to the
    intersection of these two regions associated to the regularity
    constraints. Clearly, a chirplet chain is represented by a broken
    line in the TF plane.}
\end{figure}

All smooth chirps in Eq.~(\ref{smooth_chirp}) are supported in the TF
domain $\cal{D}$, a rectangle of width $T$ and of height equal to the
Nyquist bandwidth $f_s/2$, as illustrated in Fig.  \ref{fig:cc}.  Let
$\{(t_j=j\delta_t, f_m=m\delta_f) \:;\: j=0\ldots N_t, \: m=0\ldots
N_f\}$ be a regular TF grid led on ${\cal D}$ by splitting the time
axis into $N_t$ intervals of size $\delta_t \equiv T/N_t$, and the
frequency axis into $N_f$ bins of size $\delta_f \equiv f_s/(2N_f)$.

In the following, the subscripts $j$ and $m$ designate the index of
the time interval and the frequency bin of the grid, respectively.
The index $k\in\{0,\ldots,N-1\}$ denotes the time index of a sample.

A \textit{chirplet} is a short piece of signal whose frequency varies
linearly between two successive nodes of the grid. In the time
interval $j$, we denote the time and frequency coordinates of the
chirplet extreme points by $(j,m_j)$ and $(j+1,m_{j+1})$. In the TF
plane, it is thus represented by a line joining the grid nodes $(t_j,
f_{m_j})$ and $(t_{j+1}, f_{m_{j+1}})$ (see Fig. \ref{fig:cc}).
Concretely, this means that the phase $\phi_k=\phi(t_s k)$ of a
chirplet is a quadratic function of time, as follows, for $t_j \leq
kt_s < t_{j+1}$
\begin{equation}
\label{eq:phik}
  \phi_k\equiv a_j t_{j,k}^2+b_j t_{j,k}+\theta_{j-1},
\end{equation}
where $a_j=\pi\:(f_{m_{j+1}}-f_{m_j})/\delta_t$, $b_j=2\pi\:f_{m_j}$ and
$t_{j,k}=t_s k-t_j$. 

We build the chirplet chain (CC) by enforcing chaining rules. The
frequency and phase of this chain are continuous. Clearly, the
continuity of the frequency is ensured by construction, while the
phase continuity requires that
\begin{equation}
  \theta_{j-1}=\pi\delta_t(f_{m_j}+f_{m_{j-1}})+\theta_{j-2}\,,
\end{equation}
for $j\geq 1$, and fixing $\theta_{-1}=0$. The slope of the chirplet
frequency as well as the difference between the slopes of the
frequencies of two consecutive chirplets are bounded absolutely. These
bounds are given by the two parameters $N_r'$ and $N_r''$ respectively
such that (\textit{i}) $|m_{j+1}-m_j|\leq N_r'$ and (\textit{ii})
$|m_{j+1}-2m_j+m_{j-1}|\leq N_r''$.  

Clearly, a CC is represented by a broken line in ${\cal D}$. The two
parameters $N_r'$ and $N_r''$ control the regularity of this line.
Consistently, we will refer to (\textit{i}) and (\textit{ii}) as
\textit{regularity constraints}.

The instantaneous frequency of a smooth chirp is associated to a
smooth curve in ${\cal D}$. In the same manner that broken lines are
good approximations of smooth curves, CCs are good approximations of
smooth chirps. Since CCs form a finite discrete set, they sample
\footnote{Strictly speaking, CCs don't \textit{sample} the set of
  smooth chirps since they don't belong to this set (the second
  derivative of their frequency is not defined at the boundaries of
  the grid time intervals and it is thus not bounded).} the set of smooth
chirps.  In other words, they form a \textit{template grid} of this
set.

It is important to know whether this template grid is sufficiently
tight i.e., whether for any smooth chirp, there always exists a
sufficiently close CC. The template grid tightness is controlled by
the choice of the four parameters defining the set of CCs, namely the
TF grid parameters $N_t$, $N_f$ and the regularity parameters $N_r'$
and $N_r''$. The first and preliminary step to address the tightness
question is to define a distance measuring the ``similarity'' (or
ambiguity) between two different chirps.

\subsection{Distance in the set of smooth chirps}
We follow the approach suggested in \cite{Owen:1995tm} and assume that
we ``receive'' a chirp whose phase $\phi$ is different than the
template phase $\phi^*$. We set $x_k\hat{=}s_k=A\cos(\phi_k+\theta)$,
and consider
\begin{equation}
\label{metric}
{\cal L}(\phi,\phi^*)=\frac{\ell(s;\phi)-\ell(s;\phi^*)}{\ell(s;\phi)}.
\end{equation}

Clearly, ${\cal L}$ measures the reduction factor of the ``detection
peak'' due to the mismatch between the chirp present in the data and
the chosen template. It is a relative measurement done with respect to
the ideal case where the template matches exactly the considered
chirp. In this sense, it can be interpreted as a SNR loss.

Since ${\cal L} \geq 0$ and equals 0 when $\phi=\phi^*$, it can be
interpreted as a \textit{distance} between the chirps. Note that
${\cal L}$ does not depend upon $A$.  It depends only on the phases
$\phi$ and $\phi^*$, but this dependency is difficult to perceive
intuitively from its definition in Eq.~ (\ref{metric}).

An approximated but much simpler expression can be obtained for nearby
chirp and template retaining the leading terms of a Taylor expansion
for small $\Delta_k\equiv \phi_k^*-\phi_k$. The approximation is
detailed in Appendix \ref{sec:approx_metric} and leads to the
following expression
\begin{equation}
  \label{metric_approx}
  {\cal L}(\phi,\phi^*)\approx \frac{1}{N} \sum_{k=0}^{N-1} \left(\Delta_k-\Delta\right)^2,
\end{equation}
with $\Delta=1/N\sum_{k=0}^{N-1}\Delta_k$.

Interestingly, we recognize in this expression the empirical estimate
of the variance of the phase difference $\Delta_k$. With this
definition of the distance, two chirps are ``identical'' (their
distance measured by ${\cal L}$ is zero) if and only if they have the
same phase evolution up to an additive offset.

\subsection{Is the CC grid tight ?}
In this section, we address the grid tightness problem and find the
regularity and TF grid parameters which yield a tight template grid of
CCs.  We proceed as follows: we first consider an arbitrary smooth
chirp of phase $\phi$. Then we construct a CC ``geometrically'' close
to this chirp.  We check if this CC is admissible i.e., if it
satisfies the regularity constraints. This imposes two conditions on
the parameters. Finally, we check whether it is \textit{effectively}
close to the chirp (as measured by the metric). This yields the loss
due to the approximation of the chirp by a CC in the worst case. For
tight CC grid, this loss (homogeneous to a SNR loss) has to be small
which imposes one more condition on the parameters.

\subsubsection{``Geometrically'' close CC}
Let $\phi(t)$ and $f(t)$ be the phase and frequency of an arbitrary
smooth chirp. The frequency evolution appears as a smooth curve in the
TF plane.

\begin{figure}
\includegraphics[width=\columnwidth]{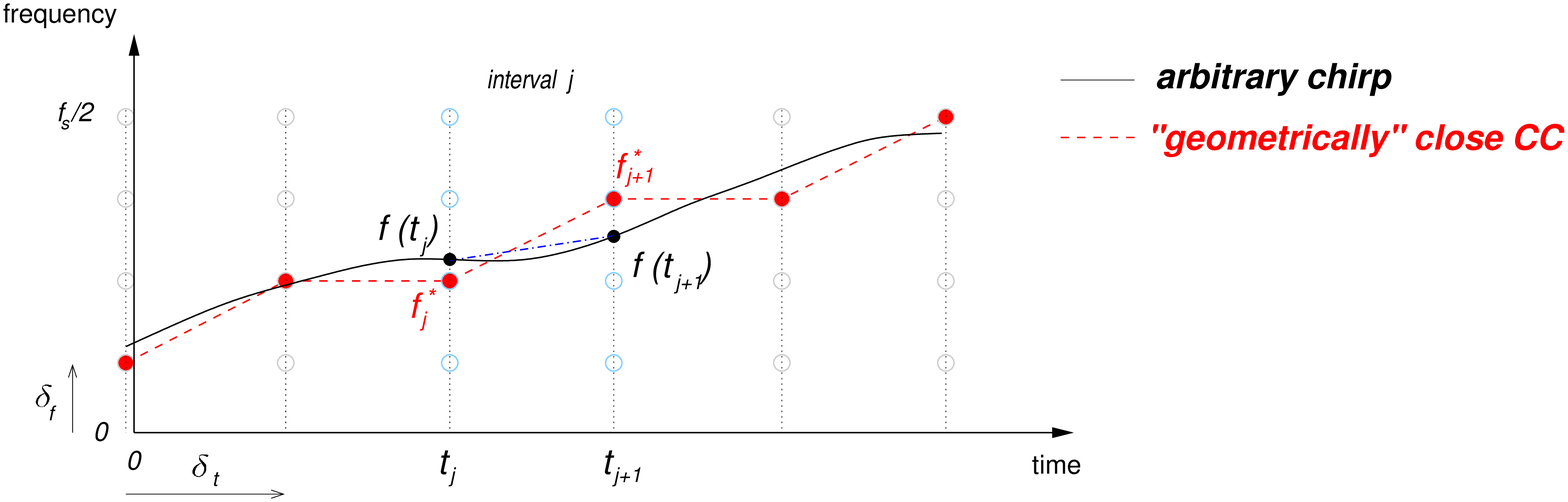}
\caption{\label{fig:close_cc}(Color online) Smooth chirp (solid curve)
  and its ``geometrically'' close CC (dotted broken line)}
\end{figure}

We construct a CC ``geometrically'' close to this chirp as follows:
for each $j=0, \ldots, N_t$, we choose the node $(t_j,f_j^*)$ of the
$j$-th column of the TF grid defined in Sec. \ref{sec:cc_def} which is
nearest to the point $(t_j,f(t_j))$. We draw the broken line by
joining these nodes (see Fig.  \ref{fig:close_cc}). The associated CC
is the ``geometrically'' close CC to the chirp under consideration and
we denote its phase $\phi^*$.

\subsubsection{Admissibility of the ``geometrically'' close CC}
For a given chirping rate limits $\dot{F}$ and $\ddot{F}$, the
``geometrically'' close CC may or may not satisfy the regularity
constraints. This depends on the regularity and TF grid parameters.
Below, we investigate this question.

\paragraph{$1^{\text{st}}$ order regularity ---}
Let us consider the chirplet of the interval $j$, we have
\begin{multline}
\label{eq:admis1}
|f^*_{j+1}-f^*_{j}|\leq |f^*_{j+1}-f(t_{j+1})|+|f(t_{j+1})-f(t_{j})|\\
+|f^*_{j}-f(t_{j})|. 
\end{multline}

Using the mean value theorem\footnote{Let the function $g(\cdot)$ be
  continuous in the open interval $(a,b)$ and differentiable in the
  closed interval $[a,b]$. The mean value theorem states that there
  exists $c$ in $(a,b)$ such that $g(b) - g(a) = \dot{g}(c) (b-a)$
  where $\dot{g}(\cdot)$ is the derivative of $g(\cdot)$.
  Consequently, we have $|g(b) - g(a)| \leq \dot{G}|b-a|$ with
  $\dot{G}=\sup_{x \in (a,b)}(\dot{g}(x))$.} (see e.g.,
\cite{Gradshteyn80}), we get $|f(t)-f(s)|\leq \dot{F} |t-s|$ from
which we deduce a bound on $|f(t_{j+1})-f(t_{j})|$.
By construction, we have $|f^*_{j}-f(t_{j})| \leq
\delta_f/2$ and this leads to
\begin{equation}
|f^*_{j+1}-f^*_{j}|\leq \delta_f + \dot{F}\delta_t. 
\end{equation}

Thus, the geometrically close CC satisfies the regularity constraint
(\textit{i}) mentioned in Sec. \ref{sec:cc_def} if
\begin{equation}
N_r'\geq \dot{F}\delta_t/\delta_f+1. 
\end{equation}

We rewrite this condition in the following form
\begin{equation}
\label{admissibilityA}
N_r'\geq 4\frac{N'}{N_t} \frac{N_f}{2N} +1,
\end{equation}
where $N'\equiv \dot{F}T^2$ is an adimensional quantity which depends
only on the fundamental characteristics of the smooth chirp model.

\paragraph{$2^{\text{nd}}$ order regularity ---}
We consider two successive chirplets in intervals $j-1$ and $j$. 
Using a similar method, the difference of their slopes can be bounded by
\begin{equation}
|f^*_{j+1}-2f^*_{j}+f^*_{j-1}|\leq |f(t_{j+1})-2f(t_j)+f(t_{j-1})|+2\delta_f. 
\end{equation}

Two consecutive applications of the mean value theorem to a function
$f(\cdot)$ which satisfies Eq.~(\ref{smooth_def}) for all $s\in[r,t]$
with $0\leq r<t\leq T$ yield the following result
\begin{multline}
|f(t)-2f(s)+f(r)|\leq \ddot{F} \left((t-s)^2+(r-s)^2\right)/2 \\+\dot{F} |t-2s+r|\,.
\end{multline}

Using $r=t_{j-1}$, $s=t_j$ and $t=t_{j+1}$, we get
\begin{equation}
|f^*_{j+1}-2f^*_{j}+f^*_{j-1}|\leq \ddot{F} \delta_t^2+2\delta_f. 
\end{equation}

Therefore, the geometrically close CC satisfies the regularity
constraint (\textit{ii}) mentioned in Sec. \ref{sec:cc_def} if
\begin{equation}
N_r''\geq \ddot{F}\delta_t^2/\delta_f+2. 
\end{equation}

We rewrite the above condition as
\begin{equation}
\label{admissibilityB}
N_r''\geq \frac{4}{3} \left(\frac{N''}{N_t}\right)^2 \frac{N_f}{2N} +2, 
\end{equation}
where $N''\equiv \sqrt{3\ddot{F}T^3}$ is an adimensional quantity
which depends only on the fundamental characteristics of the smooth
chirp model, and thus from the astrophysical input. It is related to
the maximum overall curvature of the chirp frequency or more precisely
to the largest number $\ddot{F}T^3$ of Fourier bins that the chirp
frequency can sweep, the linear trend being removed.

\subsubsection{Is the ``geometrically'' close CC effectively close?}
\label{sec:geometric_is_close}
We obtain a worst case estimate on the distance between the smooth
chirp and its ``geometrically'' close CC by bounding the variations of
their phase difference. We begin with bounding the frequency
discrepancy.
 
The starting point is the following lemma (inspired from
\cite{Arias:2004}, p.  23) obtained from the application of the mean
value theorem and some algebraic manipulations. If $f(\cdot)$
satisfies Eq.~ (\ref{smooth_def}) for all $s\in[r,t)$ then
\begin{equation}
\label{lemmaC}
\left|f(s)-\left[f(r)+\frac{s-r}{t-r}(f(t)-f(r))\right]\right|\leq \ddot{F}(t-r)^2. 
\end{equation}

However, this upper-bound can be slightly improved: the term $(t-r)^2$
over-estimates the more precise bound
$g(s)\equiv\min\{(t-s)(t-(s+r)/2),(s-r)((s+t)/2-r)\}$. (Note that $g(t)=0$
and $g(r)=0$ as expected.) In the worst case, we have $s=(t+r)/2$
and $g(s)=3/8(t-r)^2$ as opposed to
$(t-r)^2$. We include this gain in the following.

We apply the lemma in Eq.~(\ref{lemmaC}) with $t=t_{j+1}$ and $r=t_j$.
The term inside the square brackets in the left hand side of Eq.~(\ref{lemmaC})
is equal to the frequency at time $s$ of the chirplet obtained by
joining $f(t_{j-1})$ to $f(t_j)$ (see the dot-dash line in Fig.
\ref{fig:close_cc}). We denote this frequency $\tilde{f}(s)$, and then
obtain
\begin{equation}
|f(s)-\tilde{f}(s)|\leq 3\ddot{F}\delta_t^2/8. 
\end{equation}

Since $|f(s)-f^*(s)|\leq |f(s)-\tilde{f}(s)|+|\tilde{f}(s)-f^*(s)|$
and $|\tilde{f}(s)-f^*(s)|\leq \delta_f/2$, we have
\begin{equation}
\label{Deltaf}
|f(s)-f^*(s)|\leq 3\ddot{F}\delta_t^2/8 + \delta_f/2 \equiv \Delta_f. 
\end{equation}

By definition $\phi(t)-\phi(r)=2\pi\int_{r}^{t} f(s)\,ds$. Integrating
both sides of the above inequality between two successive points
$r=t_s (k-1)$ and $t=t_s k$ for $k \in \{1,\ldots, N-1\}$, we get
\begin{equation}
\label{Deltaphi}
|\Delta_k-\Delta_{k-1}|\leq 2\pi\Delta_f t_s,
\end{equation}
which constraints the variations of the phase difference
$\Delta_k=\phi^*-\phi_k^*$.

We prove in Appendix \ref{sec:const_max} that the approximated
distance ${\cal L}(\phi,\phi^*)$ as shown in Eq.
(\ref{metric_approx}) is maximum under this constraint, when
$\Delta_k=\pm 2\pi\Delta_f t_s k$.  In this case, the maximum is ${\cal
  L}(\phi,\phi^*)=(\pi\Delta_f T)^2/3 \equiv \mu'$.

We can finally state the following \textit{tight template grid}
theorem: for all smooth chirps of phase $\phi$, there exists a CC of
phase $\phi^*$ such that
\begin{equation}
\label{effective}
{\cal L}(\phi,\phi^*)\leq \mu',
\end{equation}
where $\mu'=\pi^2 T^2 (3 \ddot{F}\delta_t^2/4+\delta_f)^2/12$ is the
maximum (i.e., in the worst case) \textit{energy} SNR loss due to the
mismatch between the smooth chirp and the chosen template given by a
close CC. The corresponding maximum \textit{amplitude} SNR loss is
$\mu=1-\sqrt{1-\mu'} \approx \mu'/2$ for small $\mu'$.  Note that the
\textit{amplitude} SNR loss is linked to the \textit{minimal match}
$MM$ defined in \cite{Owen:1995tm} by the relation $MM=1-\mu$. We
express the maximum SNR loss $\mu$ in terms of the CC parameters as
\begin{equation}
\label{eq:snr}
\mu=\frac{\pi^2}{96}\left[\frac{1}{2} 
\left(\frac{N''}{N_t}\right)^2+\frac{1}{2} \left(\frac{2N}{N_f} \right)\right]^2.
\end{equation}

In principle, this loss can be made arbitrarily small by choosing
$N_t$ and $N_f$ adequately. Therefore, \textit{the grid of CC can
  sample the set of smooth chirps tightly}.

It is evident that two types of losses contribute to $\mu$. The first
one is related to the \textit{geometrical error} due to the fact that
the model is a broken line: within the time intervals of the TF grid,
the model is a straight line which cannot perfectly follow the
curvature of the smooth chirp frequency. The finer the grid along the
time axis, the smaller the time interval, the better the line fits the
smooth chirp frequency, thus reducing this error.  The other is
related to the \textit{quantization error} as we require the node of
best broken line to belong to the TF grid: there is a difference
between the best broken line we can possibly draw and the closest
(quantized) one with vertices belonging to the grid. The finer the
grid along the frequency axis, the closer the quantized line from the
original, thus reducing this error. The maximum SNR loss is the
function of these two independent parameters.

When $N_t=N''$ and $N_f=2N$, the maximum SNR loss is of order $\sim
\pi^2/96 \approx 10\%$ and the two types of errors contribute equally.
The same maximum SNR loss can be achieved with other choices for $N_t$
and $N_f$. In the next section, we propose a criterion to solve this
indetermination.

\subsection{Smallest tight CC grid}
\label{smallCC}
As elaborated in the previous section, the tight template grid theorem
gives the condition on the TF grid parameters $N_t$ and $N_f$ which
ensures that the template set (the CCs) covers all the feasible set
(the smooth chirps) with a given accuracy specified by the maximum SNR
loss. The same accuracy can be achieved with several pairs of
parameters, leading to a parameter indetermination.

For a small maximum SNR loss, the maximization of the LLR in
Eq.~(\ref{max_llr}) performed over the set of smooth chirps can be
safely replaced by a maximization over the set of CCs, i.e.,
\begin{equation}
\label{maxcc}
\ell_{\mathrm{max}}(x) \simeq \max_{\text{all CCs}}\{\ell(x;\phi)\}.
\end{equation}

The statistic $\ell_{\mathrm{max}}$ in Eq.~(\ref{maxcc}) results from
the CC which maximizes the statistic or in other words, from the
waveform of the template set which best fits the data.  Generally
speaking, when the data is noise \textit{only}, the larger the number
of (reasonably different) waveforms in the template set, the larger
the risk that one of the waveforms fits the noise and consequently,
the larger the false alarm rate.

The TF grid parameters $N_t$ and $N_f$ influence very differently the
number of CCs. The above argument suggests to select the parameters
which minimize the numbers of CCs, for a given specified maximum SNR
loss. We refer to the \textit{smallest tight CC grid} as the set of
CCs which results from this constrained optimization.

Let us first estimate the number of CCs. According to the regularity
conditions, each of the number $N_c \sim N_f(2N_r'+1)$ of possible
chirplets in a given time interval can be chained to (at most)
$2N_r''+1$ chirplets in the next time interval.  Counting CCs is then
a combinatorial problem. We have $N_c$ chirplets in the first time
interval, and $2N_r''+1$ possible choices for the $N_t-1$ successive
time intervals. Neglecting what happens at the lower and upper
boundaries of the frequency axis (i.e., near DC and Nyquist), we
obtain an upper-bound on (the logarithm of) the number $N_{cc}$ of
CCs as
\begin{equation}
\label{eq:CCno}
\ln N_{cc} \lesssim \ln(2 N_r' N_f) + (N_t-1) \ln(2 N_r'' +1).
\end{equation}

In practice, we have $N_t\gg 1$. The second term largely
dominates the right hand side and the first term can be neglected.
We thus have
$\ln N_{cc} \sim N_t \ln(2 N_r'' +1)$.

At this point, it is convenient to introduce $u\equiv N''/N_t$ and
$v\equiv 2N/N_f$ and express the \textit{smallest tight CC grid}
problem with these variables. From the regularity constraints, we have
$N_r''=4u^2/(3v)+2$.  We want to minimize the number of CCs
\begin{equation}
\ln N_{cc} \propto g(u,v) \equiv \frac{1}{u} \ln \left(\frac{8}{3}\frac{u^2}{v}+5\right),
\end{equation}
subject to a given maximum SNR loss i.e., $u^2+v=C\equiv
8\sqrt{6\mu}/\pi$.

Combining the derivatives of the objective $dg=\d_u g du + \d_v g dv$
and of the constraint $dv=-2u du$, we obtain the equation giving the
admissible point where the derivative $dg/du$ vanishes, viz.
\begin{equation}
\frac{\ln y}{y-5} + \frac{7}{4y} - \frac{3}{4}=0
\end{equation}
where we defined $y\equiv 8u^2/(3v)+5$. This equation can be solved
numerically and gives $y \approx 8.95$. Let $\alpha\equiv
u^2/v=3(y-5)/8$ be the ratio between the two errors contributing to
$\mu$. We obtain the smallest tight template grid when this ratio is
$\alpha \sim 1.48$. For a required $\mu$, we get the parameters of the
resulting grid as follows. Using the constraint, we have
$u=\sqrt{C\alpha/(1+\alpha)}$ and $v=C/(1+\alpha)$, from which we
obtain the parameters,
\begin{align}
\label{eq:smallcc}
N_t&=0.52 \: \mu^{-1/4} \: N''\,,&
N_f&=0.78 \: \mu^{-1/2} \: N.
\end{align}

Interestingly, this also implies that $N_r''= 4\alpha/3 + 2 \approx 4$ 
is a constant (i.e., does not depend on $\mu$). The last parameter $N_r'$ is
directly determined by substituting Eqs.~(\ref{eq:smallcc}) in
(\ref{admissibilityA}). 

The parameters of the smallest tight template grid may not be always
suitable in practise (see the later discussion on the implementation
and numerical contingencies in Sec. \ref{sec:cost}) but they give
interesting indications. 

At this point, it is useful to see with an
example if the proposed model and template grid sound tractable in a
realistic case.

\subsection{Toy model and CC parameters}
\label{params}
We use the inspiralling binary chirps as a toy model to check whether
the various parameters have reasonable order of magnitudes in this
physically realistic situation. We consider the Newtonian
approximation of the chirp whose frequency evolution is given by
\cite{Sathyaprakash:1991mt}
\begin{equation}
\label{newtonian}
f(t)=f_0\left(1-\frac{t-t_0}{T}\right)^{-3/8} \text{for $t<t_0+T$,}
\end{equation}
where $t_0$ denotes the arrival time. In practice, the arrival time
corresponds to the time at which the chirp enters the detector's
bandwidth i.e., when its frequency reaches the low frequency (seismic)
cut-off (denoted $f_0$) of the interferometric detectors. The $T$
defines the chirp duration, i.e.  the time taken by the chirp from the
arrival time till the binary coalescence.

The chirp duration can thus be estimated by
\begin{equation}
T\sim 1.3 \:\mathrm{s} \left(\frac{f_0}{20}\right)^{-8/3} \left(\frac{M}{50\Mo}\right)^{-5/3},
\end{equation}
where $M$ is the total mass (objects of equal masses).

In this calculation, we assume the seismic cut-off
frequency\footnote{This is the seismic cut-off frequency targetted by
  the detector Virgo.} of $20$ Hz.

We fix $\dot{F}$ and $\ddot{F}$ to the corresponding values of the
first and second derivatives of the chirp frequency, pertaining to the
last stable circular orbit (LSCO\footnote{For non-rotating stars, the
  LSCO is when the objects are at the distance $r=6 G M/c^2$.}) viz.,
\begin{align}
  f_{\rm LSCO} &\sim 88.4 \: \mathrm{Hz} \left(\frac{M}{50\Mo}\right)^{-1} \,,\\
  \dot{F}&\sim 1.33 \:\mathrm{kHz/s} \left(\frac{M}{50\Mo}\right)^{-2} \,,\\
  \ddot{F}&\sim 74 \:\mathrm{kHz/s}^2 \left(\frac{M}{50\Mo}\right)^{-3} \,.
\end{align}

We note that $T, f_{\rm LSCO}, \dot{F}$ and $\ddot{F}$ decrease with
an increasing mass. When $M$ increases, the chirp is thus shorter,
less steep and curved, and it reaches only the lower part of the
frequency band. From the above equations, we deduce that
\begin{align}
  N' &\sim 2.2 \times 10^3 \left(\frac{M}{50\Mo}\right)^{-16/3} \,,&
  N'' &\sim 698 \left(\frac{M}{50\Mo}\right)^{-4}\,.
\end{align}

The sampling frequency $f_s$ is fixed by the width of the
observational band of the GW detector, namely $f_s=2048$ Hz.  We thus
have
\begin{equation}
  N=f_s T\sim 2662 \left(\frac{M}{50\Mo}\right)^{-5/3} \,.
\end{equation}

Following Sec. \ref{smallCC} and fixing $\mu=10\%$, the smallest
tight CC grid has the following parameters for the TF grid
\begin{align}
  N_t &\sim 645 \left(\frac{M}{50\Mo}\right)^{-4},\\
  N_f &\sim 6566 \left(\frac{M}{50\Mo}\right)^{-5/3},
\end{align}
and for the regularity, we have
\begin{align}
  N_r' &\sim 17 \left(\frac{M}{50\Mo}\right)^{-4/3},&
  N_r'' &\sim 4.
\end{align}

The orders of magnitude for the various parameters appear to be
reasonable. Since these parameters don't increase with $M$, the
template grid defined with the above values remains acceptable and
tight for higher masses $M\geq 50 \Mo$.

\section{Find the best chirplet chain}
\label{sec:maxcc}
In Sec. \ref{sec:cc}, we have shown that, the SNR loss due to the use
of a CC instead of the {\it ideal} template can be made small with an
appropriate choice of parameters i.e., by making the CC grid tight.
In other words, the problem of detecting a smooth chirp is equivalent
to the one of detecting a CC as stated by Eq.~(\ref{maxcc}). The
maximization over the set of CCs -- involved in the latter case -- has the great
advantage that it can be resolved numerically.

\subsection{The exhaustive search is not feasible}
\label{sec:exhaust}
Since CCs are in finite number, an obvious maximization procedure is
to try them all and select the one which gives the maximum. To
understand whether this solution is tractable, we need to know how
many CCs are there. We consider that the search parameters $N_t$,
$N_f$, $N_r'$ and $N_r''$ are known and can be obtained from the
physical and grid tightness requirements as discussed earlier.  

We already presented an estimate of the number of CCs in Eq.
(\ref{eq:CCno}) and saw that it grows exponentially with the
number of time intervals of the TF grid. This estimate computed for
the toy model example presented in the previous section gives $\log_{10} N_{cc}
\approx 1400$.  Clearly, this number is too large for an exhaustive
search (i.e., computing $\ell$ for all possible CCs) to be carried out
in real time on existing computers.  Generally speaking, since the
number of CCs increases exponentially with $N_t$, the cost of an
exhaustive search scales exponentially with $N_t$ and thus with the
problem size $N$.

In the next section, we propose an algorithm which gives a good
estimate for the optimal CC instead of the exact solution of the
maximization problem described in Eq.~(\ref{maxcc}). However, as
opposed to the exhaustive search, the computational cost of this
algorithm scales as a polynomial of the problem size $N$.

\subsection{Near optimal search}
\label{sec:nearopt}
The maximization of $\ell(x;\phi)$ in Eq.~(\ref{maxcc}) is a combinatorial
maximization problem. The existence of an efficient solving algorithm
for such problem is related to the structural properties of the
``objective'' function to be maximized, that is, $\ell$ in the present
case.  In this Section, we show that $\ell$ can be reasonably
approximated by a path integral computed over a time-frequency
representation (TFR) of the data. The structure of the approximated
statistic allows us to perform its maximization efficiently with
dynamic programming.  The approximation goes through two stages with
an intermediate step for the reformulation of the statistic in the TF
plane.

\subsubsection{Approximation 1: for a CC, cosine and sine are almost orthogonal}
\label{sec:approx1}
As shown in Eq.~(\ref{statistic_GS2}), the statistic $\ell$ can be
expressed as
\begin{equation}
\label{statistic_GS}
\ell(x;\phi)=\frac{1}{2}\left[\left(\sum_{k=0}^{N-1} x_k \tilde{c}_k\right)^2 + 
\left(\sum_{k=0}^{N-1} x_k \tilde{s}_k\right)^2 \right],
\end{equation}
where the templates $\tilde{c}_k$ and $\tilde{s}_k$ are the orthonormalized
counterparts of the waveforms in quadrature $\cos\phi_k$ and
$\sin\phi_k$ obtained from the Gram-Schmidt procedure as given below
\be
\label{eq:csk}
\tilde{c}_k=\frac{\cos\phi_k}{\sqrt{n_c}} \quad\quad
\tilde{s}_k=\frac{n_c \sin\phi_k - n_x \cos\phi_k}{\sqrt{n_c {\cal O}}}. 
\ee
Let $\{\cos\phi_k\}$ and $\{\sin\phi_k\}$ be the vectors in $\R^N$
associated to the quadrature waveforms. As it appears in the above
expressions, these vectors are generally not orthonormal. Their
deviation from orthonormality can be quantified with two parameters,
defined by
\begin{align}
\label{eq:delep}
\delta&\equiv \frac{n_c-n_s}{n_c+n_s}\,,&\epsilon&\equiv \frac{2n_x}{n_c+n_s}\,,
\end{align}
which are related to their vector lengths $n_c$, $n_s$ and their
scalar product $n_x$. The parameter $\delta$ measures the relative
difference in the vector lengths while $\epsilon$ measures the angle
between them. The vectors are orthonormal if and only if both $\delta$
and $\epsilon$ are zero.

Intuitively, if the quadrature waveforms oscillate sufficiently, they
should be close to orthonormality and $\delta$ and $\epsilon$ are
expected to be small. This intuition is examined in details in Appendix
\ref{sec:delta_epsilon}, in which we exploit the fact that $\phi$ is
not arbitrary but it is the phase of a CC. We show that
if $\phi$ is the phase of a CC whose node frequencies are in
the bandwidth $f_l \leq f_{m_j} \leq f_s/2-f_l$ for all $j$ with
\begin{equation}
  f_l \approx 2.5 \delta_f \sqrt{N'}\left(\frac{N_f}{N}\right)^{3/2} \left(\frac{0.1}{\eta}\right)^{1/2}\,,
\end{equation}
then $|\delta|\lesssim \eta$ and $|\epsilon|\lesssim \eta$.

In the following, we assume that this condition is satisfied. This
imposes the CC frequency not to approach arbitrarily close to the DC
nor to the Nyquist frequencies. Since the amplitude of the
instrumental noise of GW interferometers diverges rapidly when going
close to DC, it is not expected to detect GWs at low frequencies.
Therefore the reduction of the bandwidth in the low frequency region
should not be a problem as long as $f_l$ remains small. We will check
later with examples that the reduction of the useful bandwidth is
indeed sufficiently small.

Using Eqs.~(\ref{eq:csk}) and (\ref{eq:delep}), we can write
$\tilde{c}_k$ and $\tilde{s}_k$ in terms of $\delta$ and $\epsilon$ as
\bea 
\tilde{c}_k &=& \left(\frac{2}{N (1 +\delta)}\right)^{1/2}~\cos \phi_k \, ,\\
\tilde{s}_k &=& \left(\frac{2}{N (1 +\delta)}\right)^{1/2}~
\frac{(1+\delta)\sin \phi_k - \epsilon \cos \phi_k}{\sqrt{1 - (\delta^2 +\epsilon^2)}}\,,
\eea 
noting that $n_c=N(1+\delta)/2$, $n_s=N(1-\delta)/2$ and
$n_x=N\epsilon/2$.

Inserting this expression in Eq.~(\ref{statistic_GS}) and taking the
limit for small $\delta$ and $\epsilon$, we find that $\ell(x;\phi)
\rightarrow \hat{\ell}(x;\phi) \equiv (x_c^2 + x_s^2)/N$, the
reminder $R(\delta,\epsilon) \equiv \hat{\ell}(x;\phi)-\ell(x;\phi)$
being given by
\be
\label{eq:Rem}
R(\delta,\epsilon)=\frac{1}{N} \frac{\delta (x_c^2 - x_s^2) + 2 \epsilon x_c x_s -(\delta^2+\epsilon^2)(x_c^2+x_s^2)}{1-(\delta^2+\epsilon^2)}.
\ee

Considering that we have $\delta^2+\epsilon^2\leq \eta^2$ (see Appendix
\ref{sec:delta_epsilon}) and
\begin{align}
\label{eq:rel}
\frac{|x_c^2 - x_s^2|}{x_c^2 + x_s^2} &\leq 1,  &\frac{|2x_c x_s|}{x_c^2 + x_s^2} &\leq 1,
\end{align}
the relative error can be bounded as
\be
\label{eq:Rem2}
\frac{|R(\delta,\epsilon)|}{\hat{\ell}(x;\phi)} \leq \frac{2\eta-\eta^2}{1-\eta^2} \approx 2\eta,
\ee
for small $\eta$.

Provided a good choice of $\eta$ (and checking the consequences on
$f_l$), this approximation error can be made small. We can safely
replace $\ell$ by $\hat{\ell}$ which we express as the following
complex sum~:
\begin{equation}
\label{approx1}
\hat{\ell}(x;\phi)=\frac{1}{N}\left|\sum_{k=0}^{N-1} x_k \exp (i\phi_k) \right|^2.
\end{equation}

\subsubsection{Go to time-frequency: Moyal}
The expression of $\hat{\ell}$ in Eq.~(\ref{approx1}) computes the
canonical hermitian scalar product between the data and a complex
template waveform. While Parseval's formula allows an equivalent
formulation of this scalar product in the frequency domain,
\textit{Moyal's formula} does the same in the TF
domain, provided the use of a unitary TFR. One
such TFR is the \textit{discrete Wigner-Ville} (WV) distribution
defined in \cite{Eric:2005a} and given by
\begin{equation}
  \label{dwv}
  w_x(n,m) \equiv \sum_{k=-k_n}^{k_n} x_{p_{n,k}}  x^*_{q_{n,k}} \, e^{-2\pi i mk/(2N)},
\end{equation}
with $k_n\equiv \min\{ 2n, 2N-1-2n \}$, $p_{n,k} \equiv \lfloor n+k/2
\rfloor$ and $q_{n,k} \equiv \lfloor n-k/2 \rfloor$ where $\lfloor 
\cdot \rfloor$ gives the integer part. The arguments of $w_x$ are the
time index $n$ and the frequency index $m$ which correspond in
physical units, to the time $t_n=t_s n$ and the frequency is $f_m=f_s
m/(2N)$ for $0\leq m\leq N$ and $f_m=f_s (N-m)/(2N)$ for $N+1\leq
m\leq 2N-1$. Thus, the frequency axis gets sampled at twice the usual
rate (as performed by the FFT). The WV distribution is associated with
a particular sampling of the TF plane. As discussed later in
Sec.~\ref{sec:cost}, this leads to some restrictions on the TF grid
used for defining CCs.

Let $\{x_k\}$ and $\{y_k\}$ be two time series. Moyal's formula states that
\cite{Eric:2005a}
\begin{equation}
\label{moyal3}
\left|\sum_{k=0}^{N-1} x_k \, y^*_k \right|^2=
\frac{1}{2N} \sum_{n=0}^{N-1} \sum_{m=0}^{2N-1}w_x(n,m)\, w_y(n,m).
\end{equation}

Using Eqs. (\ref{approx1}) and (\ref{moyal3}), we rewrite $\hat{\ell}$
as the inner-product of two TFRs namely, the WV of the data $w_x$ and
the \textit{template WV} $w_e$ which is the WV of complex template
waveform $e_k \equiv \exp i\phi_k$,
\begin{equation}
\label{moyal}
\hat{\ell}(x;\phi)=\frac{1}{2N^2}\sum_{n=0}^{N-1}\sum_{m=0}^{2N-1} w_x(n,m) w_e(n,m).
\end{equation}

Chirp signals are easily modelled and described in the TF plane.
Qualitatively, we expect that the TFR of a chirp signal have large
values essentially in the vicinity of a curve corresponding to their
instantaneous frequency and vanishes elsewhere. The template WV $w_e$
being the TFR of a chirp, it shares these characteristics. In the
following section, we make use of this feature to simplify the
statistic.

\subsubsection{Approximation 2: the WV of a CC is almost Dirac}
\label{sec:approx2}
With continuous time and frequency variables, it is well-known that
(\cite{Flandrin:1999}, p. 130 and also 217) the WV of a \textit{linear
  chirp} (i.e., a chirp whose frequency is a linear function of time)
is a Dirac distribution along the TF line associated to the chirp
frequency.

We assume that this remains reasonably true for discrete time and
frequency and when the chirp is non-linear (and in particular, when it
is a CC). More precisely, we consider that we have
\begin{equation}
\label{dwv_model}
w_e(n,m)\approx 2N \:\delta(m - m_n),
\end{equation}
where $m_n=[ 2T\:f_n ]$ where $[\cdot]$ denotes the nearest integer.

Here, $f_n$ is the instantaneous frequency of the (possibly
non-linear) chirp. Eq.~(\ref{dwv_model}) dissembles two approximations
which we explain now.

\begin{figure}
  \centerline{\includegraphics[width=\columnwidth]{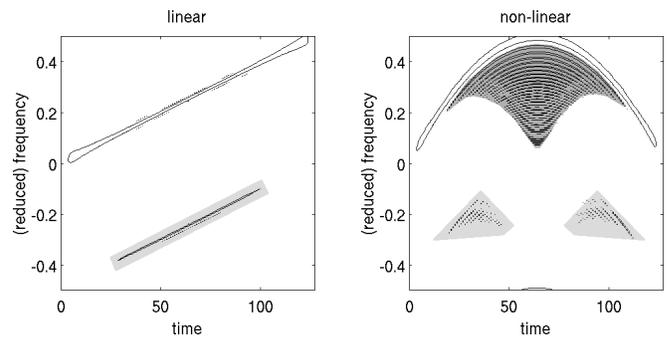}}
  \caption{\label{fig:chirp}\textbf{Discrete Wigner-Ville of two chirp
      signals} --- The signals are normalized to unit $\ell_2$ norm
    and we show the contour at the level 1/8.  \textit{left}: when the
    chirp frequency is a linear function of time, its WV is almost
    Dirac along the corresponding TF line in the TF half plane
    associated to positive frequencies. For the negative frequencies,
    the WV distribution exhibits \textit{aliasing terms} (we highlight
    them by a background in light gray) which we neglect in the
    simplified model in Eq.~({\protect \ref{dwv_model}}).
    \textit{right}: when the chirp frequency is not linear (here, it
    is a parabolic chirp), \textit{interference terms} appear
    (evidenced by a dark gray background).  Their contribution are
    also disregarded in the simplified model.  Note that the WV of the
    non-linear chirp chosen for this illustration do present aliasing
    terms (with a light gray background), but they have a smaller
    amplitude than in the linear case.}
\end{figure}

For discrete time and frequency, the discrete WV of a linear chirp can
be calculated analytically \cite{Eric:2005a}. For the positive
frequencies i.e., for $0 \leq m \leq N$, the model in Eq.
(\ref{dwv_model}) is an acceptable approximation of the exact result,
as illustrated in Fig.~\ref{fig:chirp}. However, there is a
significant difference in the negative frequencies i.e., for $N+1 \leq
m \leq 2N-1$. In this region, the discrete WV exhibits
\textit{aliasing terms} (clearly seen in the left panel of Fig.
\ref{fig:chirp}) which are closely related to the unitarity property
of the WV. In \cite{Eric:2005a}, the aliasing terms are shown to be
oscillating terms (switching signs) with smaller amplitude than the
preponderant terms modelled by Eq.~(\ref{dwv_model}). We can then
expect their contribution to the summation in Eq.~(\ref{moyal}) to be
negligible.

It is well known \cite{Flandrin:1999} that \textit{interference terms}
appear when computing (both continuous and discrete) WVs of non-linear
chirps. They can be related to the quadratic nature of this
distribution (see \cite{Flandrin:1999} for a detailed analysis of the
nature and geometry of these interferences). Interference terms change
sign rapidly (see Fig.~\ref{fig:chirp}, right panel) and can be
neglected for the same argument invoked for aliasing terms.

Inserting Eq.~(\ref{dwv_model}) into Eq.~(\ref{moyal}), we get the
following approximation of $\hat{\ell}$~:
\begin{equation}
\label{eq:pathint}
\tilde{\ell}(x;\phi) = \frac{1}{N} \sum_{n=0}^{N-1} w_x(n,m_n).
\end{equation}

We see that this statistic results from the integral of the WV of the
data along the TF path determined by the CC frequency $f_n$. In other
words, this integral is the area under this TF path. We refer to this
quantity as the \textit{path length} \footnote{If we see the WV as a
  Lebesgue measure (although this is an misuse of language since the
  WV can take negative values), the integral in Eq.~(\ref{eq:pathint})
  effectively defines a path length.}.

With this approximation, the maximization of the statistic in Eq.
(\ref{maxcc}) amounts to finding the path giving the largest integral,
or the longest path. Efficient methods exist for longest path problems
\cite{Bertsekas:2000}.  These methods exploit the structural
properties of path length (or integral) measurement, in particular,
{\it additivity}. The length of this entire path can be measured by
splitting the path and summing the length of its constituent parts.
Thanks to this property, the maximization problem can be decomposed
into a recursive series of small problems, each of them being solvable
in polynomial time.  This is the main principle of \textit{dynamic
  programming} (DP), which we describe in the next Section.

We remind the reader that, contrarily to the new statistic
$\tilde{\ell}$, the exact statistic $\ell$ is not additive. DP
\textit{cannot} be applied to maximize $\ell$.

\subsubsection{Maximization with dynamic programming}
DP is a classical method \cite{Bertsekas:2000} for solving
combinatorial optimization problems. As explained in the previous
section, the idea is to decompose the problem into smaller ones that
can easily be solved. In our context, the natural decomposition is
given by the tiling of the time axis into chirplet intervals i.e.,
$t_j \leq t_s n < t_{j+1}$ or equivalently $jb \leq n \leq (j+1)b-1$
where $b\equiv\delta_t/t_s$ is the number of samples in an interval. The
overall path integral is equal to the sum of the integrals computed in
each chirplet interval marked with an superscript index as follows
\begin{align}
\tilde{\ell}(x;\phi)&=\sum_{j=0}^{N_t-1} \tilde{\ell}^j (x;\phi)\\
\label{pathint_chirplet}
\text{with~} \tilde{\ell}^j(x;\phi)&\equiv \frac{1}{N}\sum_{n=jb}^{(j+1)b-1}w_x(n,m_n^j),
\end{align}
where $m_n^j=[2N f_n^j]$ and the frequency $f_n^j$
follows the line joining the grid points $(t_j,f_{m_j})$ and
$(t_{j+1},f_{m_{j+1}})$. We also denote with a subscript index $j$,
the path integral up to interval $j$, viz.
\begin{equation}
\tilde{\ell}_j(x;\phi)=\sum_{j'=0}^{j} \tilde{\ell}^{j'} (x;\phi).
\end{equation}

\begin{figure}
  \centerline{\includegraphics[width=\columnwidth]{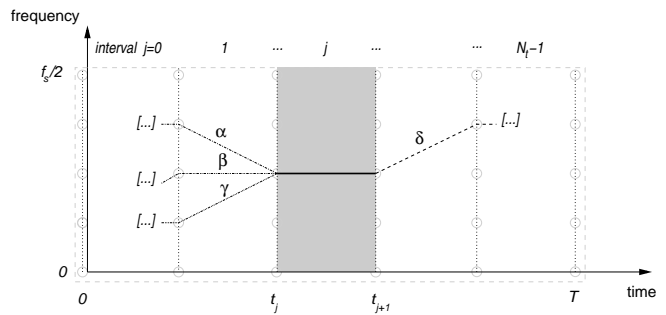}}
  \caption{\label{fig:dp}\textbf{Principle of optimality of DP}}
\end{figure}

DP relies on the \textit{principle of optimality}. We elaborate this
principle with the help of Fig. \ref{fig:dp}. We consider the chirplet
in time interval $j$.  In a chain passing through this chirplet, the
regularity constraints limit the choice of preceding chirplets in the
time interval $j-1$. We suppose that there are only three such
chirplets; namely $\alpha$, $\beta$ and $\gamma$.

Now, consider the time interval $j-1$. We assume that we know the
chain passing through the chirplet $z$ ($z$ being either $\alpha,
\beta$ or $\gamma$) and giving the largest path integral summed up to
the interval $j-1$. We denote this quantity by
$\tilde{\ell}^{(z)}_{j-1}$.  (In this discussion, the chirplet and its
associated CC are designated by the same label).

We compute the path integral contribution in $j$-th interval for the
considered chirplet, and add the result to $\tilde{\ell}^{(z)}_{j-1}$
to obtain $\tilde{\ell}^{(z)}_{j}$ for all the three paths $z= \alpha,
\beta$ and $\gamma$.

We mark with $(\star)$ the optimal chain associated to the
\textit{global} maximum of $\tilde{\ell}$ (i.e., summing from interval
$0$ to $N_t-1$) which we denote $\tilde{\ell}^{(\star)}$. We further
assume that this optimal chain $(\star)$ follows $(\alpha)$ up to interval
$j-1$, continues following the considered chirplet in interval $j$ and
proceeds to the last interval $j=N_t-1$ with some chain $(\delta)$, hence
$\tilde{\ell}^{(\star)}=\tilde{\ell}^{(\alpha)}_j+\tilde{\ell}^{(\delta)}$ where
$\tilde{\ell}^{(\delta)}$ denotes the contribution of the chain $(\delta)$.

The principle of optimality states that the optimal chain $(\star)$
has the largest path integral $\tilde{\ell}^{(\star)}_{j-1}$ at
interval $j-1$ as compared to {\it all} the other chains passing by
the same chirplet in interval $j$. In particular, this means that
$\tilde{\ell}^{(\star)}_{j-1}=\tilde{\ell}^{(\alpha)}_{j-1}$ is larger that
$\tilde{\ell}^{(\beta)}_{j-1}$ and $\tilde{\ell}^{(\gamma)}_{j-1}$.

\noindent {\it Proof by contradiction}: Let us assume that
$\tilde{\ell}^{(\beta)}_{j-1}>\tilde{\ell}^{(\alpha)}_{j-1}$. We construct the
chain $(\triangle)$ formed by $(\beta)$, the considered chirplet in
interval $j$ and the chain $(\delta)$. This CC is admissible. By
construction, its path integral
$\tilde{\ell}^{(\triangle)}=\tilde{\ell}^{(\beta)}_j+\tilde{\ell}^{(\delta)}$ is
larger than $\tilde{\ell}^{(\star)}$. Therefore, the chain $(\star)$
is not optimal which contradicts our hypothesis --- QED.

We apply this principle recursively starting from interval $j=0$ and
incrementing. For each chirplet interval and for all $N_c$ chirplets
of interval $j$, we keep only the CC maximizing the path integral up
to this point and we discard the others. This procedure ``prunes the
combinatorial tree'' and avoids to consider useless candidates before
going to the next interval.

When the recursion reaches the last interval $N_t-1$, we end up with a
number $N_c$ of CCs ending with a different last chirplet and having
the maximum path integral among all chains with the same last
chirplet. Finally, within these ``short-listed'' candidates, we select
the chain with the largest $\tilde{\ell}$ which is the \textit{global
  maximum}.

\subsubsection{Numerical contingencies and computational
  cost}
\label{sec:cost}
To summarize, we started with the initial problem in Eq.~(\ref{maxcc})
of finding the CC with the largest statistic. We rephrased this
problem (using approximations) into a {\it longest path} problem in
the TF plane. Here, \textit{path} refers to the TF curve followed by
the frequency of the CC, and the \textit{length} is given by the
integral of the WV of the data along the path. The maximization of the
path length over the set of CCs can be performed efficiently using DP.
The resulting algorithm is tractable numerically as shown by the
estimate obtained in the second part of this section.

The definition of the CCs does not comprehend the fact that we only
have access to discretized versions of the data and of their
associated TF domain, denoted $\cal{D}$ in Sec.~\ref{sec:cc_def}. We
begin this section by a discussion on these aspects.

\paragraph{Discretization issues ---}
On one hand, the definition of the set of CCs relies on a TF grid
sampling the \textit{continuous} TF domain ${\cal D}$. Theoretically,
this grid can be refined arbitrarily. On the other hand, the search
operates effectively using the \textit{discretized} version of ${\cal
  D}$, resulting from the sampling associated to the WV. This fixes a
maximum TF resolution which cannot be surpassed.

It is useless to increase the resolution of the TF grid used for
defining CCs beyond the one defined by WV. The WV divides the time
axis into $N$ intervals and the frequency axis into $N$
bins\footnote{It is possible to modify slightly the definition of WV
  in Eq.~({\protect \ref{dwv}}) to get a finer sampling of the
  frequency axis and keep unitarity. We reserve this possibility for
  later investigations.}.  Consequently, we have the following
limitations, $N_t\leq N$ and $N_f\leq N$. Furthermore, in order to
have time intervals (resp. frequency bins) of equal size , the TF grid
parameters $N_t$ (resp. $N_f$) must be divisors of $N$.

All these requirements limit the choice of $N_t$ and $N_f$.  It may
happen that the parameters of smallest tight CC grid are not suitable
because of that. Note that in the case we consider, we are generally
led to adopt the finest resolution for the frequency axis i.e.,
$N_f=N$.

\paragraph{Estimate of the computational cost ---}
We estimate of the computational cost by counting the floating point
operations for all the primary subparts in the course of the
procedure. The computation of the WV of the data involves $N$ FFTs
with time base $2N_f$ \cite{Eric:2005a}, such that the cost of
this part is about $5 N N_f \log_2 N_f$ (assuming a standard
implementation with RADIX-2).

The number of operations required by DP is better estimated by
grouping them by types, rather than by a sequential assessment.  The
path integral $\tilde{\ell}_j$ in Eq.~(\ref{pathint_chirplet}) is
computed (with $b$ additions) only once for each $N_c$ chirplets of
all $N_t$ intervals, with a corresponding cost equal to $N_c N$. 

For each of the $N_c$ chirplets in each interval, the algorithm
selects among the (at most) $2N_r''+1$ possibly connected paths. This
procedure is repeated $N_t-1$ times, and thus requires $\sim N_t N_c
(2N_r''+1)$ operations.

Knowing that the number of chirplets is $N_c\approx (2N_r'+1)N_f$, the
overall cost $C$ thus scales with
\be
\label{eq:cost}
C \propto 5 N N_f \log_2 N_f+[N+(2N_r''+1)N_t](2N_r'+1)N_f,
\ee
which is a polynomial of the problem size.

\section{Applications}
\label{sec:test}
In this section, the proposed method is evaluated with several
numerical tests and compared with two other TF based algorithms for
the detection of unmodelled chirps, namely the Signal Track Search
(STS) \cite{Anderson:2000yy} and Time-Frequency Clusters (TFC)
\cite{Sylvestre:2002ih}. The simulation code \footnote{Freely
  distributed scripts are available at
  \url{http://www.obs-nice.fr/ecm} for reproducing all the
  illustrations presented here.} of these tests uses the
implementation of these algorithms provided by
\cite{lsc_algor_librar}.  We first give a brief presentation of STS
and TFC.

\subsection{Existing algorithms}
\label{sec:others}

\subsubsection{Signal Track Search}
We have seen earlier that the TFR of a chirp signal can be essentially
described in the TF plane as a regular alignment of large values
forming ``ridges'' along the instantaneous frequency evolution. The
STS uses this observation as a heuristic basis: detecting chirps
amounts to finding ridges in a TFR.

In practise, the algorithm extracts the ridges from the WV
distribution\footnote{In \cite{Anderson:2000yy}, the authors use the
  standard definition of the discrete WV originally proposed by
  Claasen-Mecklenbr\"auker (see \cite{Eric:2005a} for a definition and
  a detailed discussion). This definition differs from the one
  presented in Sec.  \ref{sec:approx2}. In particular, it does not
  satisfy unitarity.} of the data. Because of the presence of noise,
image processing techniques are required to get a good ridge
extraction.  The authors chose an algorithm which is normally used for
road extraction from aerial images. This algorithm is based on the
fact that a ridge is a locus of points having a maximum curvature (as
measured by the second derivative) in the transverse direction and a
small gradient along the longitudinal direction. A hysteresis
thresholding procedure is applied over the second derivative of the WV
(smoothed by a low pass filter) to detect TF points which suffice the
above condition, and to grow iteratively chains of TF points from
these ridge precursors. In \cite{Anderson:2000yy}, the ridge length
(number of TF points in a ridge) is then employed as the detection
statistic.  However, we don't use this definition here, but we rather
consider the one given by the largest path integral computed along the
detected ridges. We observed that this variation outperforms the
original definition of STS.

\subsubsection{TFClusters}
TFClusters is initially thought to detect short oscillatory transients
(and not specifically chirps). The TFR of such transient is sparse
i.e., the TF contents is essentially described by few components of
large amplitude. The basic idea of TFClusters is that, for reasonable
SNR, the amplitude of the transient components is larger than the
noise.

This motivates the thresholding of the TFR of the data, given by the
spectrogram (modulus square of short-time Fourier transform), to
retain the TF points with the largest values. A clustering algorithm
is used to group the selected points. ``Significant'' clusters are
chosen whose cardinals are greater than a threshold.
``Insignificant'' clusters are merged iteratively if they are
sufficiently close to eventually form significant clusters. The
statistic is then chosen to be the maximum sum of the TF powers over
the clusters in the resulting list of significant clusters.

\subsubsection{Discussion}
It is important to stress a major difference between STS and
TFClusters and the proposed method.  By construction, the formers work
well provided that the signal ``stands above'' the noise somewhere in
the TF plane.  If we define a local SNR in the TF plane (by computing
at a given TF point, the squared difference of the mean values of the
TFR under the hypotheses $H_0$ and $H_1$ divided by its variance under
$H_0$), then this is equivalent to say that the \textit{local} SNR has
to be large at least for some TF points.  However, just like the
standard matched filtering, a detection with the best CC algorithm
requires the \textit{global} SNR (obtained by summing the local SNRs
for all TF points) to be large.  Clearly, this is a less stringent
condition.

TF path integration is a central ingredient of the best CC search.
This idea is also used for other methods developed for the detection
of other GW sources, for instance, for inspiralling binaries, we can
cite \cite{chassande-mottin99,morvidone03:_time} and for the periodic
GW sources, the Hough transform \cite{krishnan04} and the stack slide
searches \cite{brady00}.

Several distinctions must be stressed. First, the TF representation we
use here (discrete WV) satisfies a specific and crucial property,
namely unitarity. This allows us to link the final statistic to the
quadratic matched filtering. TF representations based on short-time
Fourier or wavelet bases used by the above methods are not unitary.
Second, other methods require a precise model of the TF path (relying
on the astrophysical source modelling) as opposed to our method. For
the problem adressed here i.e., the detection of unmodeled chirps, we
have shown that CCs can be treated as an effective finite template
grid. We could then imagine to apply one of the above methods and
integrate along the entire set of TF paths associated to CCs. This is
however computationally impossible because of the too large number of
CCs, as already discussed in Sec.~\ref{sec:exhaust}.

\subsection{Newtonian chirps: illustrations and benchmark}
For the illustration of the best CC search, we use the Newtonian chirp
signal introduced in Sec. \ref{params}. We recall that the frequency
of such chirp is a power law given by Eq.~(\ref{newtonian}). Normally,
the Newtonian chirp also includes a prescribed evolution of the chirp
amplitude. However, for simplicity and better match with our initial
model, we decide not to take this into account and keep the chirp
envelope to a constant.

\begin{figure}
  \centerline{\includegraphics[width=\columnwidth]{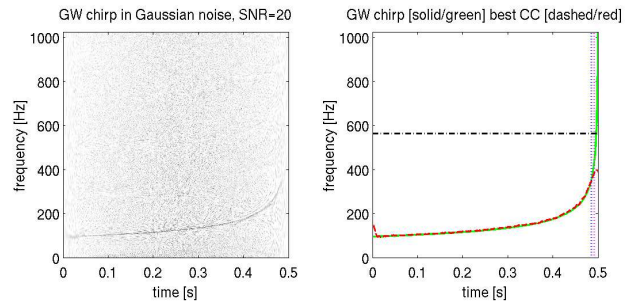}}
  \caption{\label{fig:figure0a}(Color online) \textbf{Newtonian chirp
      in white Gaussian noise} --- \textit{left}: WV distribution of
    the signal. Only positive contributions are displayed (negative
    ones are set to zero) with a grey-scaled color map going from
    white (minimum i.e, zero) to black (maximum). \textit{right}: the
    best CC in dashed/red closely matches the actual instantaneous
    frequency in solid/green in the region where the regularity
    constraints are satisfied. We indicate the instant when the chirp 
    reaches the
    chirping rate limits with the dotted vertical lines and the frequency
    at LSCO with dashed-dotted horizontal line.}
\end{figure}

The Newtonian chirp is completely defined by the total mass $M$ of the
binary (if we assume that the objects have equal masses) and its
initial frequency $f_0$. Fig. \ref{fig:figure0a} presents an example
of a typical Newtonian chirp signal, where we set $M=7.3 \Mo$ and
$f_0=96$ Hz. The chirp duration is $T=0.5$ s.  We fix the sampling
frequency to $f_s=2048$ Hz (therefore, the number of samples is $N=T
f_s=1024$). A white Gaussian noise of unit variance is added to the
signal.

Within the GW literature, it is customary to define the SNR through
matched filtering (assuming the initial phase is known \textit{a
  priori}). We follow this definition which gives in the present case,
\begin{equation}
  \rho^2 \equiv \sum_{k=0}^{N-1} s_k^2 \approx A^2 N/2.
\end{equation}

We note that, with this definition we have $\rho^2 = 2 \ell (s;\phi)$
(the factor of 2 accounts for the unknown initial phase).

We choose to scale the chirp amplitude to a SNR $\rho=20$.

We apply the best CC search to this signal with the following search
parameters. We arbitrarily fix the chirping rate limits to be
$\dot{F}=8192$ Hz/s and $\ddot{F}=917.5$ kHz/s$^2$. These values are
quite smaller than the ones expected at the LSCO (see Sec.
\ref{params}) but the time instant when theses limits are reached is
close (few tenths of milliseconds before) to the LSCO. In Fig.
\ref{fig:figure0a}, the time instants when the chirp (the solid line
on the right panel) reaches the chirping rate limits (with dotted
vertical lines) and when the binary system reaches LSCO (with
dashed-dotted horizontal line) are indicated. We fix the frequency
axis sampling to the finest accessible resolution i.e., $N_f=N=1024$.
Similarly, we choose the smallest possible chirplet size with
$N_t=N/2$. The rest of the parameters are derived from the regularity
constraints.  In this respect, it is useful to calculate the
adimensional characteristics of the problem i.e., $N'=2048$ and
$N''=586.6$.  The resulting parameters are $N_r'=9$ and $N_r''=3$,
which gives a maximum SNR loss $\mu \approx .28$.

We recall that the best CC search relies on the approximation of the
optimal statistic by a complex sum presented in
Sec.~\ref{sec:approx1}. The parameter $\eta$ controls the relative
precision of this approximation. From the results of
Sec.~\ref{sec:approx1} and the above general chirp specifications, the
approximation holds with a precision $\eta=0.14$ in a frequency
bandwidth $[f_l,f_s/2-f_l]$ with $f_l= 96$ Hz which coincides (at
least for the low frequencies, which are most important) with the
frequency support of the present chirp.

Fig. \ref{fig:figure0a} presents the result of the best CC search with
the above choice of parameters. The best CC closely matches the actual
instantaneous frequency in the region where the regularity constraints
are satisfied.

An example is obviously not sufficient to evaluate the method
thoroughly. Receiver operating characteristics (ROC) gives a
systematic assessment of the performance. The ROC of a given
statistic $l$ is the diagram giving the detection probability
$P_d(l_0) \equiv \P(l \geq l_0|H_1)$ versus the false alarm
probability $P_{fa}(l_0)\equiv\P(l \geq l_0|H_0)$ at a given
SNR and for all thresholds $l_0$.

For this exercise, due to computing limitations, we prefer
short signals with a small number of samples
$N$. We choose a Newtonian chirp with
total mass $M=11 \Mo$ and initial frequency $f_0=96$ Hz which has a
short duration $T \sim 250$ ms. Choosing the sampling frequency $f_s=1024$
Hz, we have $N=256$ samples.  White Gaussian noise is added to the
signal and the amplitude is scaled such that the SNR is $\rho=10$.

We fix the chirping rate limits to $\dot{F}=8.192$ kHz/s and
$\ddot{F}=1.05$ MHz/s$^2$. Like the above example, these limits are
reached at a time instant close to the LSCO. We choose the finest TF
grid parameters $N_t=128$ and $N_f=256$, and the regularity parameters
$N_r'=9$, $N_r''=4$. The resulting CC grid is tight with $\mu\approx
0.4$.

\begin{figure}
  \centerline{\includegraphics[width=\columnwidth]{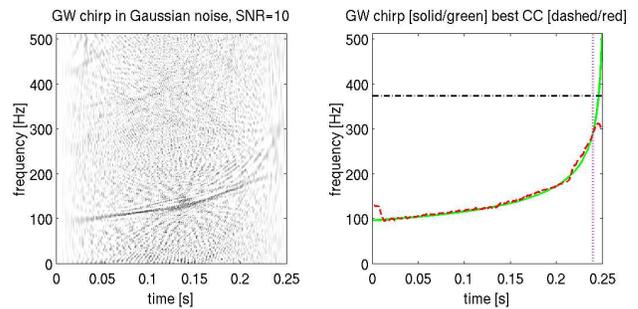}}
  \caption{\label{fig:figure0b} (Color online) \textbf{Newtonian chirp
      in white Gaussian noise} --- \textit{left}: WV distribution of
    the signal (displayed similarly as in Fig. {\protect
      \ref{fig:figure0a}}). \textit{right}: actual chirp frequency in
    solid/green and best CC in dashed/red. We indicate where the chirp
    reaches the chirping rate limits with dotted vertical lines and
    the frequency at LSCO with dashed-dotted horizontal line.}
\end{figure}

\begin{figure}
  \centerline{\includegraphics[width=\columnwidth]{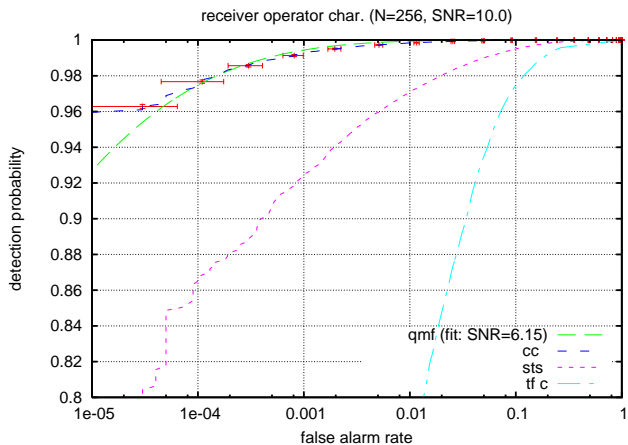}}
  \caption{\label{fig:figure1} (Color online) \textbf{Newtonian chirp
      in white Gaussian noise} --- Comparison of ROCs of the best CC search
    (dashed/blue, with error bars in solid/red) with STS (dotted/magenta)
    and TFC (dashed-dotted/cyan). The computation
    of each ROCs is perfomed over $2\times 10^5$ trials (half for the false
    alarm probability and half for the detection probability). The
    diagram also includes the ROC of the clairvoyant quadrature
    matched filter (bold dashed/green) shown here with the SNR
    $\rho_c=6.15$ adjusted to reasonably fit the ROC of the best CC
    search.}
\end{figure}

Concerning STS\footnote{Following the notations of
  \cite{Anderson:2000yy}, the size of the Gaussian kernel of the
  pre-smoothing filter is fixed to $\sigma=2$. The low and high
  thresholds of the hysteresis are set to 3.3/pixel$^2$ and 10/pixel$^2$
  resp.}  and TFClusters\footnote{The TFR is given by the short-time
  Fourier transform computed over non-overlapping blocks of 16 samples
  (i.e., intervals of $\approx 7.8$ ms). The frequency axis is tiled
  into 32 bins (i.e., a resolution of $32$ Hz). We use the nominal
  values given in \cite{Sylvestre:2002ih} for the rest of the
  parameters namely $p=0.1$, $\sigma=5$,
  $\mathbf{\delta}=[0,0,0,0,0,0,2,3,4,4]$ and $\alpha=0.25$.}, we set
their free parameters empirically using the recommendations available
in the references, without a precise fine-tuning. Fig. \ref{fig:figure0b}
displays a single trial and Fig.
\ref{fig:figure1} presents the ROCs of the three methods presented
previously. We see that the best CC search outperforms the two others
as expected.

Here, we wish to add few remarks regarding the comparison between the
best CC search and STS. The improvement in the ROC of the best CC with
respect to STS has two origins. First, the use of a unitary discrete
WV instead of the standard WV helps in increasing the detection
probability by few percent. The unitarity preserves the power in TF
plane and hence improves the efficiency. Second, the major part of the
improvement comes from the TF pattern search procedure. As explained
in Sec. \ref{sec:others}, the use of a global search criterion instead
of a local one is a crucial ingredient.

It is interesting to compare these ROCs with what could optimally
achieve an imaginary observer which knows in advance the targeted
chirp. Since this \textit{clairvoyant} observer knows the chirp phase
exactly, he can apply the optimal statistic i.e., the quadrature
matched filter obtained in Appendix \ref{sec:llr}. The ROCs of the
quadrature matched filter can be obtained analytically (under Gaussian
noise hypotheses). The false alarm and detection probabilities are
given respectively by \cite{johnson70:_distr_statis}:
\begin{align}
  P_{fa}(l_0)&=\exp(-l_0),\\
  P_d(l_0)&=1-\exp(-\rho_c^2/2) \sum_{n=0}^{+\infty}
  \frac{(\rho_c^2/2)^n}{n!} I_{l_0}(n+1)
\end{align}
where $I_y(x)\equiv 1/\Gamma(x) \int_0^y e^{-u}u^{x-1}\:du$ is the
incomplete Gamma function. 

This ROC depends only on one parameter, namely the SNR $\rho_c$.  The
ROC curve of clairvoyant statistic with $\rho_c=\rho$ provides an
absolute upper bound on the detection probability. Obviously, having
in hand all the information makes a very large difference with respect
to the case where we only know that the incoming GW is a smooth chirp.
The detection probability of the clairvoyant statistic is very close
to 1 over the entire range of value chosen for the false alarm rate.
This is why we don't show this curve. It is more interesting to
compare the performances of the various methods with the ones of the
clairvoyant observer for SNRs $\rho_c < \rho$. More precisely, we
adjust $\rho_c$ in such a way that the resulting curve matches
reasonably well the ROC of the best CC search in the region of
interest i.e., for false alarm probabilities in the range $10^{-5}$ to
$10^{-4}$. Since the SNR is inversely proportional to the distance of
the GW source, the ratio of the actual SNR to the best-fit value
$\rho/\rho_c$ gives the reduction factor of the sight distance with
respect to the ideal (and non accessible) situation where we have at
our disposal all the information about the chirp we want to detect.
We include the fitted clairvoyant ROC in Fig. \ref{fig:figure1}. The
ratio in the sight distance can be estimated $\sim 10/6.15\approx
1.6$.

\subsection{Random CCs and robustness}
While benchmarks based on Newtonian chirps are satisfactory for a
comparison of several detection methods in a nominal situation, they
don't provide a test for the robustness i.e., a measurement of
their ability to detect reliably a large class of different chirps.

In this section, we present ROC curves computed using \textit{random
  CCs}. Random CCs are generated by chaining chirplets randomly chosen
in a range specified by regularity constraints. Therefore, the
frequency of a random CC follows a kind of random walk in the
TF plane. We generate a new random CC for each trial made
to estimate the detection probability.

The detection of random CCs is obviously much more difficult than the
detection of a single chirp. It is an effective test of the method
robustness. No classical approaches (e.g., based on banks of
quadrature matched filters as for inspiralling binary chirps) can be
applied successfully in this case.

We assume the same general characteristics of the Newtonian chirp used
in the first example in the previous section, namely $T=0.5$ s,
$f_s=2048$ Hz, thus $N=1024$ samples, $\dot{F}=8192$ Hz/s and
$\ddot{F}=917.5$ kHz/s$^2$. We already computed satisfactory search
parameters for this set-up. Therefore, they remain unchanged
($N_t=512$, $N_f=1024$, $N_r'=9$ and $N_r''=3$). The random CCs are
generated on the same basis, but with a time interval slightly larger,
the regularity parameters being increased accordingly i.e., $N_t=64$,
$N_f=1024$, $N_r'=65$ and $N_r''=57$. We use an additive white
Gaussian noise.

\begin{figure}
  \centerline{\includegraphics[width=\columnwidth]{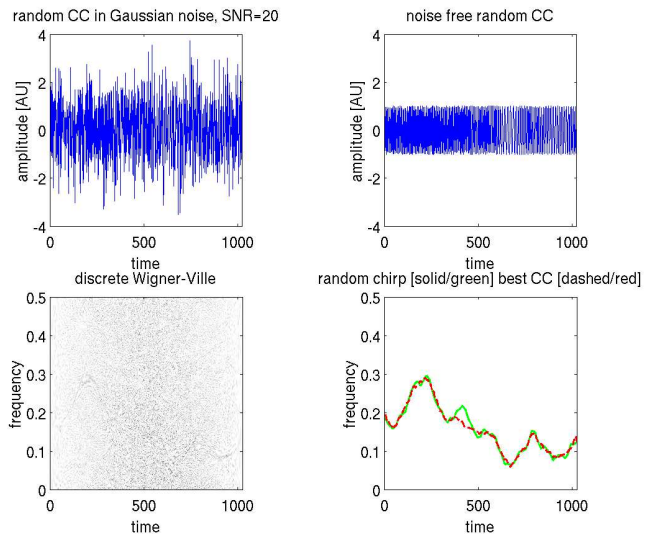}}
  \caption{\label{fig:figure0c} (Color online) \textbf{Random CC in
      white Gaussian noise} --- In these plots, we arbitrarily set
    $f_s=1$. \textit{top -- left}: example of a random CC in white
    Gaussian noise. \textit{top -- right}: noise free random CC. ---
    \textit{bottom -- left}: WV distribution of the signal (displayed
    similarly as in Fig.  {\protect \ref{fig:figure0a}}).
    \textit{bottom -- right}: actual chirp frequency in solid/green
    and best CC in dashed/red. It is worthwhile to note that, although
    the best CC can lose track for some time because of noise
    fluctuations, it is able to recover the exact TF path.}
\end{figure}

\begin{figure}
  \centerline{\includegraphics[width=\columnwidth]{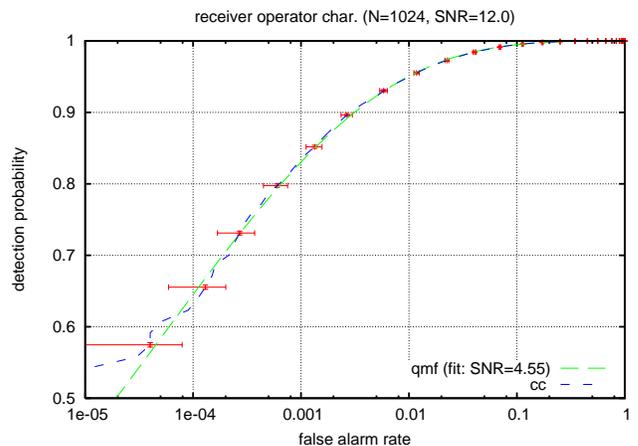}}
  \caption{\label{fig:figure2} (Color online) \textbf{Random CC in
      white Gaussian noise} --- This diagram displays the ROC of the
    best CC search (dashed/blue, obtained from $2\times 10^5$ trials)
    compared with the analytical ROC of the clairvoyant matched filter
    (bold dashed/green) with the SNR $\rho_c=4.55$, adjusted to
    produce a reasonable fit.}
\end{figure}

Fig. \ref{fig:figure0c} presents an example of such signal (with SNR
$\rho=20$) and the result of the application of the best CC search.
Fig.  \ref{fig:figure2} displays the ROC curve of the best CC search
(with SNR $\rho=12$) along with the one of the clairvoyant quadrature
matched filter adjusted to an adequate SNR.  We estimate a loss in the
sight distance with respect to the clairvoyant case to be a factor of
$\sim 2.6$.  Best CC search ``sees'' to distances comparable to (in
the sense, with a reduction factor less than one order of magnitude)
what classical methods achieve in other GW detection problems.

The computational cost of this search as estimated by
Eq.~(\ref{eq:cost}) is about $142$ millions of floating points
operations for one block of duration $T=0.5 s$. Assuming $10\%$
overlap between successive blocks, real-time processing can be achieved
with a computing power of 2.8 Gflops which is less than what a single
standard workstation can handle today.

\section{Concluding remarks}
Smooth chirps define a general model of ``nearly physical'' GW chirps.
Chirplet chains -- chains of linear chirplets -- allow the design of
tight template grids for the detection of smooth chirps. The optimal
detection requires these grids to be searched thoroughly to find the
template which best matches with the data. Although the shear large
number of templates prevents the use of an exhaustive search,
near-optimal detection can be performed with the time-frequency based
procedure presented here. Its originality lies in the clear link
established between the optimal statistic and the proposed search
algorithm as opposed to with other approaches. In particular, it
justifies the choice of a specific time-frequency representation (the
unitary discrete WV) and pattern search algorithm (dynamic
programming). We have evaluated that best CC search is computationally
tractable for detection of typical GW chirps.

It is important to emphasize several features which makes the proposed
method attractive in practise. First, the free parameters (the chirp
duration $T$ and the chirping rate limits $\dot{F}$ and $\ddot{F}$)
are few and directly related to physical characteristics.  Second, the
principle ``\textit{He who can do more can do less}'' applies here:
smooth chirps is a very general class of chirps. This model, and thus
the search algorithm can be easily modified and adapted to incorporate
additional astrophysical information.  For instance, it is easy to
search only chirps with an increasing (or decreasing) frequency.  One
may also want a more stringent constraint on the chirping rate at low
frequencies than at high frequencies. The inclusion in the algorithm
of a dependency of the chirping rate limit upon the frequency is
straightforward.  This leaves the possibility of a compromise between
efficiency (since the restriction of the set of admissible waveforms
due to additional constraints reduces the false alarm rate) and
robustness, depending on the quantity and reliability of the
information available on a specific GW source.  Third and finally, it
is simple to restrict the search to chirps starting and/or finishing
at given time-frequency location. This feature could be used for
partially known chirps whose waveforms is known only on a part of the
total duration. Those signals could be detected with a hybrid approach
combining a standard matched filtering where the waveform model is
available, and best chirplet chain search for the rest.

\begin{acknowledgments}
  This work has benefited from many discussions and exchanges of
  ideas. For this reason, we would like to thank Sanjeev Dhurandhar,
  Patrick Flandrin, J\'er\^ome Idier, Ekatherina Karatsuba and Bruno
  Torresani. We gratefully acknowledge ICTP (Trieste, Italy),
  Observatoire de la C\^ote d'Azur (France) and EGO (Cascina, Italy)
  for their support and hospitality.
\end{acknowledgments}

\appendix
\section{Maximizing LLR $\Lambda(\cdot)$ over the initial phase $\varphi_0$}
\label{sec:llr}
In this appendix, we maximize the statistic
$\Lambda(x;\{\hat{A},\varphi_0,t_0,\phi(\cdot)\})$ over the initial phase
$\varphi_0$. We recall that
\begin{equation}
\Lambda(x;\{\hat{A},\varphi_0,t_0,\phi(\cdot)\})=\frac{1}{2{\cal N}}\left(\sum_{k=0}^{N-1} x_k \bar{s}_k\right)^2,
\label{Eq:lambda1}
\end{equation}
where ${\cal N}\equiv \sum_{k=0}^{N-1}\bar{s}^2_k$ is the norm of
$\bar{s}_k=\cos(\phi_k+\varphi_0)$. To keep the notations simple, we
don't mention all parameters explicitly and set
$\Lambda(x;\varphi_0)=\Lambda(x;\{\hat{A},\varphi_0,t_0,\phi(\cdot)\})$.

In the literature concerning the detection of inspiralling binaries of
compact objects \cite{Sathyaprakash:1991mt,Owen:1995tm}, this
maximization is usually performed assuming that ${\cal N}$ is
independent of $\varphi_0$. This assumption is correct when the two
quadratures $\cos\phi_k$ and $\sin\phi_k$, viewed as vectors of
$\R^N$, are orthonormal (i.e., orthogonal and of same norms). In this
case, we have $n_c=n_s=N/2$ and $n_x=0$ where $n_c$, $n_s$ and $n_x$
are the norms and cross-products of the quadratures as defined in Eqs.
(\ref{normal_var}). Inserting this into
\begin{equation} 
{\cal N}=n_c\cos^2\varphi_0-n_x \sin(2\varphi_0)+n_s\sin^2\varphi_0,
\end{equation}
we conclude that ${\cal N}=N/2$ is a constant.

However, for general phase evolution, the quadrature waveforms are not
necessarily orthonormal. This is approximately true when the chirp
oscillates sufficiently rapidly during many cycles (e.g., for
inspiralling binaries of a small mass). Since we are considering
chirps with an arbitrary phase and of relatively short duration, such
assumption is not realistic and we opt for the general case keeping
the dependency of ${\cal N}$ upon $\varphi_0$.

Expanding $\bar{s}_k$ in terms of two quadratures and rewriting
Eq.~(\ref{Eq:lambda1}), we get
\begin{equation}
\label{llr_a}
\Lambda(x;\varphi_0)=\frac{(x_c\cos \varphi_0 - x_s \sin\varphi_0)^2}
{2(n_c\cos^2 \varphi_0 - n_x \sin(2\varphi_0) +n_s\sin^2 \varphi_0)},
\end{equation}
where $x_c$ and $x_s$ are the cross-correlation of the data with
$\cos\phi_k$ and $\sin\phi_k$ as defined in Eq.~(\ref{cos_sin}).

To proceed with the maximization, we first examine the special case
where the quadratic waveforms are linearly dependent i.e., $\cos\phi_k
\propto \sin\phi_k$ for all $k$. This implies that we are in the
degenerate case where $\phi_k=\varphi_0$ is constant.  Introducing the
two angles $\varphi=\arg(x_c+ix_s)$ and
$\eta=\arg(\sqrt{n_c}+i\sqrt{n_s})$, we can rewrite Eq.~(\ref{llr_a})
as
\begin{equation}
  \Lambda(x;\varphi_0)=\frac{1}{2}\frac{x_c^2+x_s^2}{n_c+n_s}\frac{\cos^2(\varphi+\varphi_0)}{\cos^2(\eta+\varphi_0)}.
\end{equation}

The proportionality of the quadrature waveforms implies that
$\sqrt{n_c}x_s \pm \sqrt{n_s}x_c=0$ which gives $\sin(\eta -
\varphi)=0$, and hence $\eta= \varphi+\pi\Z$. We conclude that
$\Lambda(x;\varphi_0)$ remains constant for all $\varphi_0$
and is equal to the statistic given by
\begin{equation}
\ell(x;t_0,\phi)=\frac{x_c^2+x_s^2}{2N}.
\end{equation}

In the non degenerate case, we compute the derivative of the statistic
as given in Eq.~(\ref{llr_a}) w.r.t. $\varphi_0$. Its numerator turns
out to be a second order polynomial of $\tan \varphi_0$. The root
associated to the local maximum is
\begin{equation}
  \hat{\varphi}_0=\tan^{-1} \left( \frac{x_s n_c - x_c n_x}{n_x x_s - n_s x_c} \right)
\end{equation}
which gives the ML estimator of the initial phase.

Inserting this expression in Eq.~(\ref{llr_a}) yields
\begin{equation}
\ell(x;t_0,\phi)\equiv\Lambda(x;\hat{\varphi}_0)
=\frac{n_s x_c^2 - 2 n_x x_c x_s + n_c x_s^2}{2{\cal O}}\,,
\end{equation}
where ${\cal O}\equiv n_c n_s-n_x^2 >0$.

We can re-express this statistic as
\begin{equation}
\label{statistic_GS2}
\ell(x;t_0,\phi)=\frac{1}{2}\left[\left(\sum_{k=0}^{N-1} x_k \tilde{c}_k\right)^2 + 
\left(\sum_{k=0}^{N-1} x_k \tilde{s}_k\right)^2 \right],
\end{equation}
where $\tilde{c}_k$ and $\tilde{s}_k$ are the orthonormalized
counterparts of the waveforms in quadrature $\cos\phi_k$ and
$\sin\phi_k$ obtained from the Gram-Schmidt procedure as given below
\be
\tilde{c}_k=\frac{\cos\phi_k}{\sqrt{n_c}} \quad\quad
\tilde{s}_k=\frac{n_c \sin\phi_k - n_x \cos\phi_k}{\sqrt{n_c {\cal O}}},
\ee
and referred to as templates of $\phi$.

In practise, this orthonormalization is indeed performed for the
detection of inspiralling binaries (see \cite{bala96}, p.~3046)
and is justified with heuristic arguments. The derivation shows that
it results directly from the maximization of the LLR.
 
\section{Taylor approximation of the distance between chirps}
\label{sec:approx_metric}

In this appendix, we detail the approximation of the statistic
$\ell(s;\phi^*)$ with $s_k = A\cos(\phi_k+\varphi_0)$ and assuming
that the template phase $\phi^*$ is close to the phase $\phi$ of the
signal $s$ present in the data. We start from the following Taylor
expansion of $\ell(s;\phi^*)$ for small $\Delta_k\equiv\phi_k^*-\phi_k$
\begin{multline}
\label{taylor}
\ell(s;\phi^*)=\ell(s;\phi)+\sum_{k=0}^{N-1} \d_k\ell|_{\phi^*=\phi}\Delta_k\\
+\frac{1}{2} \sum_{k,l=0}^{N-1} \d^2_{kl}\ell|_{\phi^*=\phi}\Delta_k \Delta_l + \ldots,
\end{multline} 
where the partial derivatives $\d_k\equiv\d/\d \phi^*_k$ and
$\d^2_{kl}\equiv\d^2/\d \phi^*_k\d \phi^*_l$ are taken with respect to
the samples of the template phase $\phi^*$. Next, we examine this
expansion term by term and obtain analytical expressions as a function
of the phase samples $\{\phi_k\}$ and $\{\phi^*_k\}$.

\subsection{First derivative: local extremum}
From Eq.~(\ref{eq:stat1}), we write the statistic $\ell$ as the ratio
$\ell=n/d$. The numerator is $n=n_s x_c^2 - 2 n_x x_c x_s + n_c x_s^2$
and the denominator is $d=2(n_c n_s-n_x^2)$. We thus have $\d_k
\ell=(\d_k n-\ell \d_k d)/d$.

We get the following general expressions of the derivative of the
numerator \footnote{Here, we adopt the precedence rule $\d_k ab=(\d_k
  a)b$.}
\begin{multline}
\d_k n=\d_k n_s x_c^2 + n_s 2x_c\d_k x_c -2(\d_k n_x x_c x_s + n_x \d_k x_c x_s
\\
+n_x x_c \d_k x_s) + \d_k n_c x_s^2 + n_c 2x_s \d_k x_s,
\end{multline}
and of the denominator
\begin{equation}
\d_k d=2(\d_k n_c n_s + n_c \d_k n_s -2 n_x \d_k n_x).
\end{equation}

We insert $s_k=A\cos(\phi_k+\varphi_0)$ and work out each of their
component term. At the match (when $\phi^*=\phi$), we get
\begin{gather}
\d_k n_s|_{\phi^*=\phi}= \sin 2\phi_k\,, \qquad \d_k n_x|_{\phi^*=\phi} =\cos 2\phi_k \,,\\
\d_k n_c|_{\phi^*=\phi}= - \sin 2 \phi_k = -\d_k n_s|_{\phi^*=\phi} \,,\\
\d_k x_s|_{\phi^*=\phi}= A \cos \phi_k \cos(\phi_k+\varphi_0) \,.\\
\d_k x_c|_{\phi^*=\phi}=-A \sin \phi_k \cos(\phi_k+\varphi_0) \,,
\end{gather}

Combining all the above expressions, the
derivative $\d_k n$ can be factorized, yielding
\begin{equation}
\d_k n|_{\phi^*=\phi}=\ell(s;\phi) \d_k d|_{\phi^*=\phi} \,,
\end{equation}
where $\d_k d|_{\phi^*=\phi}=2(n_c-n_s)\sin 2\phi_k-4n_x\cos 2\phi_k$
and $\ell(s;\phi)=A^2(n_c\cos^2\varphi_0-n_x\sin
2\varphi_0+n_s\sin^2\varphi_0)/2$. In conclusion, the first derivative
$\d_k\ell|_{\phi^*=\phi}=0$ vanishes at $\phi^*=\phi$ which is
thus a local extremum.

Using the parameters $\epsilon= 2n_x/N$ and $\delta= (n_c-n_s)/N$ as
defined later in Sec. \ref{sec:nearopt} (and also discussed in
Appendix \ref{sec:delta_epsilon}), the statistic and the denominator
at the match can be expressed as functions of $\delta$ and $\epsilon$
as
\begin{align}
\label{l}
\ell(s;\phi)&=\frac{A^2 N}{4}(1+\delta\cos 2\varphi_0-\epsilon\sin 2\varphi_0)\\
d|_{\phi^*=\phi}&=\frac{N^2}{2}(1-\delta^2-\epsilon^2).
\end{align}

\subsection{Second derivative and distance}
We show in the previous subsection that the first derivative at the
match $\d_k\ell|_{\phi^*=\phi}$ vanishes.  Consequently, the second
derivative at the match can be expressed simply in terms of the second
derivatives of the numerator and denominator at the match, namely
\be
\d^2_{kl}\ell|_{\phi^*=\phi} = \left[ \frac {\d^2_{kl} n-\ell(s;\phi^*)
\d^2_{kl} d}{d}\right]_{\phi^*=\phi}\,.
\ee

We obtain the following general expressions for the second derivatives of the denominator
\begin{multline}
\label{d2_denominator}
\d^2_{kl} d=2[\d^2_{kl} n_c n_s + \d_k n_c \d_l n_s + \d_l n_c \d_k n_s + n_c \d^2_{kl} n_s \\
- 2 (\d_k n_x \d_l n_x + n_x \d^2_{kl} n_x)].
\end{multline}
and of the numerator
\begin{widetext}
\begin{multline}
\label{d2_numerator}
\d^2_{kl} n=
\d^2_{kl}n_s x_c^2 + 2 (\d_k n_s x_c\d_l x_c + \d_l n_s x_c \d_k x_c + n_s \d_k x_c \d_l x_c + n_s x_c\d^2_{kl} x_c)\\
-2(\d^2_{kl} n_x x_c x_s + \d_k n_x \d_l x_c x_s + \d_k n_x x_c \d_l x_s 
  + \d_l n_x \d_k x_c x_s + \d_l n_x x_c \d_k x_s \\
+ n_x \d_k x_c \d_l x_s + n_x \d_l x_c \d_k x_s + n_x x_s \d^2_{kl} x_c + n_x x_c \d^2_{kl} x_s)\\
+ \d^2_{kl}n_c x_s^2 + 2 (\d_k n_c x_s\d_l x_s + \d_l n_c x_s \d_k x_s + n_c \d_k x_s \d_l x_s + n_c x_s\d^2_{kl} x_s),
\end{multline}
\end{widetext}

Similarly to the first derivative, we insert the expression of the
signal $s_k=A\cos(\phi_k+\varphi_0)$ and evaluate each of the
component terms of the above expressions. We have to distinguish two
cases i.e., the non-diagonal cross terms of the Hessian matrix when
$k\neq l$ and the diagonal ones when $k=l$.

\subsubsection{Cross terms, $k\neq l$}
When $k\neq l$, the above Eqs. (\ref{d2_denominator}) and
(\ref{d2_numerator}) are significantly simplified because all the
second order cross derivatives are zeros (namely
$\d^2_{kl}n_c=\d^2_{kl}n_x=\d^2_{kl}n_s=0$ and $\d^2_{kl}
x_c=\d^2_{kl} x_s=0$). We get
\begin{equation}
\d^2_{kl} d|_{\phi^*=\phi}=-4\cos 2(\phi_k-\phi_l),
\end{equation}
and combined with Eq. (\ref{l}),
\begin{widetext}
\begin{multline}
[\d^2_{kl} n-\ell(s;\phi^*)\d^2_{kl} d]_{\phi^*=\phi}=
\frac{A^2 N}{4}\left[1+\cos 2(\phi_k-\phi_l)-\cos 2(\phi_k+\varphi_0)-\cos 2(\phi_l+\varphi_0)\right.\\
+\epsilon\,(\sin 2(\phi_k+\phi_l+\varphi_0)-\sin 2\phi_k-\sin 2\phi_l-\sin 2\varphi_0)\\
\left.+\delta\,(\cos 2(\phi_k+\phi_l+\varphi_0)-\cos 2\phi_k-\cos 2\phi_l+\cos 2\varphi_0)\right]
\end{multline}
\end{widetext}

In Appendix \ref{sec:delta_epsilon}, we discuss the range of values
taken by $\epsilon$ and $\delta$ depending on the phase $\phi$. We
show that these parameters are small $\epsilon, \delta \ll 1$ if the
phase $\phi$ is a CC whose frequency does not come close to DC nor the
Nyquist frequency. We assume that this remains true in the more
general case, when $\phi$ is the phase of a smooth chirp. We retain
the leading term (of order 0 in $\epsilon$ and $\delta$) and get the
following approximation
\begin{equation}
\label{X}
\d^2_{kl} \ell|_{\phi^*=\phi} \equiv X_{kl} = \frac{A^2}{2N}\left((1-\hat{c}_k)(1-\hat{c}_l)+\hat{s}_k\hat{s}_l\right)
\end{equation}
where $\hat{c}_k=\cos 2(\phi_k+\varphi_0)$ and $\hat{s}_k=\sin
2(\phi_k+\varphi_0)$.

\subsubsection{Auto terms, $k=l$}
We consider the case where $k=l$. Now, the second order derivatives do
not vanish. In fact, we have
\begin{gather}
\d^2_k n_c|_{\phi^*=\phi}=-2\cos 2\phi_k \qquad \d^2_k n_x|_{\phi^*=\phi} =-2\sin 2\phi_k\\
\d^2_k n_s|_{\phi^*=\phi}=2\cos 2 \phi_k\\
\d^2_k x_c|_{\phi^*=\phi}=-A/2(\cos(2\phi_k+\varphi_0)+\cos\varphi_0)\\
\d^2_k x_s|_{\phi^*=\phi}=-A/2(\sin(2\phi_k+\varphi_0)-\sin\varphi_0).
\end{gather}

The consequence is an additional term $D_{kl}$ to the second order
derivative of the statistic $\d^2_{kl}
\ell|_{\phi^*=\phi}=X_{kl}+D_{kl}$.  With a direct calculation, we
obtain its exact expression (no approximation needed):
\begin{equation}
\label{D}
D_{kl}=\frac{A^2}{2}(-1+\hat{c}_k)\delta_{kl}.
\end{equation}
where the Kronecker symbol is $\delta_{kl}=0$ for $k\neq l$ and $1$
for $k=l$.

\subsection{Approximated distance}
From Eqs. (\ref{X}), (\ref{D}) and assuming that $|\epsilon|, |\delta|
\ll 1$, we have $\ell(s;\phi)\approx A^2 N/4$. The distance defined in
Eq.~ (\ref{metric}) can thus be written as
\begin{multline}
{\cal L}(\phi,\phi^*)\approx \frac{1}{N} \sum_{k=0}^{N-1} (1-\hat{c}_k)\Delta_k^2 
- \left(\frac{1}{N} \sum_{k=0}^{N-1} (1+\hat{c}_k) \Delta_k\right)^2 \\
- \left(\frac{1}{N} \sum_{k=0}^{N-1} \hat{s}_k \Delta_k\right)^2.
\end{multline}

Considering that $\Delta_k$ and $\Delta_k^2$ are slowly varying with
respect to $\hat{c_k}$ and $\hat{s}_k$, we argue that, similarly to
what is discussed in Appendix \ref{sec:delta_epsilon}, the
positive and negative terms compensate when making the following sums
$\sum_k \hat{c}_k \Delta_k$, $\sum_k \hat{s}_k \Delta_k$ and $\sum_k
\hat{c}_k \Delta_k^2$. We neglect the small residual, which leads to
the final approximation of the distance in Eq.~(\ref{metric_approx}).

\section{Constrained maximization of the distance}
\label{sec:const_max}
We rewrite the constrained maximization problem described in Sec.
\ref{sec:geometric_is_close} of the distance in Eq.
(\ref{metric_approx}) under the constraint in Eq.~(\ref{Deltaphi})
with simpler notations. We relate them to the initial problem at the
end of this Appendix.

Let $\{r_k\}$ a series of $N$ real numbers. We want to maximize the
empirical variance $V(r)$ expressed by 
\begin{equation}
V(r) =\frac{1}{N} \sum_{k=0}^{N-1} r_k^2 - \left( \frac{1}{N} \sum_{k=0}^{N-1} r_k \right)^2,
\end{equation}
under the constraint that the increments $u_k\equiv r_k-r_{k-1}$ are
absolutely bounded by some constant $U>0$ i.e., $|u_k|\leq U$ for
$k>0$.

The empirical variance $V(r)$ is invariant by the addition of an
arbitrary constant $C$: let $r_k=y_k+C$, for all $k$, then
$V(r)=V(y)$. We can thus assume with no loss of generality that
$r_0=0$ (i.e., choose $C=-y_0$). Therefore, we have
$r_k=\sum_{j=1}^{k} u_j$ for $k>0$.

We want to maximize the convex function $V$ in the set of feasible
solution described by $\{r_k\}$ which is a polyhedron of $\R^N$. From
a classical theorem of convex analysis (see \cite{Bertsekas:2003}, p.
187), we conclude that $V$ reaches its maximum at one of the extreme
points of this polyhedron. The extreme points are the points where
the increments are either $u_k=+U$ or $u_k=-U$. There are $2^{N-1}$
extreme points and we need to identify the one which maximizes the
convex function.

Let us rewrite the empirical variance $V(r)$ as a function of $u_k$.  We
leave the ``auto-terms'' $u_k^2$ aside (for all extreme points, the
auto-terms are equal to $U^2$ independently of the sign of $u_k$.
Their contribution is thus unimportant for the identification of the
maximum) and concentrate on ``cross-terms'' (i.e., terms in $u_j
u_k$).  A direct calculation leads to
\begin{equation}
V(r)=V_a+\sum_{j=1}^{N-2}\sum_{k=j+1}^{N-1} c_{jk} u_j u_k,
\end{equation}
where $c_{jk}=2j(N-k)/N^2$ and $V_a$ is the contribution due to the
auto-terms.

Since all $c_{jk}>0$, the maximum of $V$ is reached when all $u_k$
have the same signs, that is when $u_k$ are all identically $+U$ or
$-U$. Therefore, the empirical variance is maximum when $r_k=\pm k U$
and in this case $V(r)=U^2(N^2-1)/12$.

We recall that the distance between the smooth chirps is well approximated 
by the empirical variance of the phase discrepancy [see Eq.\ref{metric_approx}].
We apply this result to the original maximization problem by
setting $r_k\hat{=}\Delta_k$ and $U\hat{=}2\pi\Delta_f t_s$ as given
in Eq.~(\ref{Deltaphi}).

\section{Bounding $\delta$ and $\epsilon$ of a CC}
\label{sec:delta_epsilon}
The simplification of the statistic in Sec.~\ref{sec:nearopt} is
closely related to the orthogonality and length difference of the
vectors $\vec{c}\equiv \{\cos\phi_k, k=0,\ldots,N-1\}$ and
$\vec{s}\equiv \{\sin\phi_k, k=0,\ldots,N-1\}$ of $\R^N$.

Noting that their norms and scalar-product are respectively given by
$n_c=\langle \vec{c},\vec{c} \rangle$, $n_s=\langle \vec{s},\vec{s}
\rangle$ and $n_x=\langle \vec{c},\vec{s} \rangle$ as defined in
Eq.~(\ref{normal_var}), the departure from ``orthonormality'' of
$\vec{c}$ and $\vec{s}$ can be quantified by the two parameters 
\begin{align}
\delta&=\frac{n_c-n_s}{n_c+n_s} & 
\epsilon&=\frac{2n_x}{n_c+n_s}.
\end{align}

The parameter $\delta$ measures the relative difference of the lengths
of $\vec{c}$ and $\vec{s}$ while $\epsilon$ is related to the cosine
of the angle between the two vectors.

When the vectors $\vec{c}$ and $\vec{s}$ are orthonormal i.e.,
orthogonal and of same lengths, both $\delta$ and $\epsilon$ are zero.
By continuity, for nearly orthonormal vectors, $\delta$ and $\epsilon$
are then expected to be small.  Intuitively, this should be true for
vectors with oscillating components like $\vec{c}$ and $\vec{s}$.
Indeed, $\epsilon$ and $\delta$ can be rewritten in the form of
oscillating sums, namely
\begin{align}
\label{eq:delep1}
\delta&=\frac{1}{N}\sum_{k=0}^{N-1}\cos 2\phi_k&
\epsilon&=\frac{1}{N}\sum_{k=0}^{N-1}\sin 2\phi_k.
\end{align}

The positive and negative contributions cancel in the summation, and
thus leaves a small residual. In this appendix, we go beyond this
intuitive rationale when the phase $\phi$ is a CC as defined in
Eq.~(\ref{eq:phik}) and give a systematic investigation of the maximum
value taken by $\delta$ and $\epsilon$.

Eq.~(\ref{eq:delep1}) motivates us to combine $\delta$ and $\epsilon$
is the following complex sum $S$ 
\be
\label{eq:S}
S\equiv \delta+i\epsilon=\frac{1}{N}\sum_{k=0}^{N-1}\exp i2\phi_k.
\ee

Bounding the modulus of $S$ is equivalent to bounding $\delta$ and
$\epsilon$. Analytic number theory provides a large number of results
concerning exponential sums like $S$, for improving upon the trivial
bound $|S| \leq 1$. We use one of these, namely the Kuzmin-Landau
lemma, see \cite{graham91:_van} p. 7. We present a proof of this lemma
pertaining to the present case where the phase $\phi$ is a CC.

The proof can be summarized as follows. A change of variables is
introduced which allows us to put a bound on the modulus of $S$ by a
sum of the finite difference of complex variables.  These new
variables appear to be collinear in the complex plane. The sum of the
modulus of their difference is thus equal to the distance between the
extremes.  The final bound on $|S|$ is then obtained by combining this
property with the explicit expression of the phase of the CC, provided
a constraint on the lower and higher frequencies reached by the CC.

Let us define for $1 \leq k \leq N-1$, the following variables
\begin{align}
\label{eq:xi_zeta}
d_k &\equiv 2(\phi_k-\phi_{k-1})&
\zeta_k &\equiv \frac{1}{1-\exp(i d_k)}.
\end{align}

We perform the above change of variables in the sum $S$ using the relation
\be
\label{eq:rec}
\exp(i2\phi_k)=[\exp(i2\phi_k)-\exp(i2\phi_{k+1})] \zeta_{k+1}\,,
\ee
and we get
\begin{multline}
NS =\zeta_1 \exp(i2\phi_0) + 
\sum_{k=1}^{N-2} (\zeta_{k+1}-\zeta_k) \exp(i2\phi_k) + \\
(1-\zeta_{N-1}) \exp(i2\phi_{N-1}). \label{eq:a}
\end{multline}

By taking the modulus on both side, we obtain the following bound,
\begin{multline}
N|S| \leq |\zeta_1| + |1-\zeta_{N-1}| + 
\sum_{j=0}^{N_t-1}\sum_{k=jb+1}^{(j+1)b-1} |\zeta_{k+1} - \zeta_k| +\\
\sum_{j=1}^{N_t-2} |\zeta_{jb+1} - \zeta_{jb}|.
\label{eq:c}
\end{multline}
where we split the sum in Eq.~(\ref{eq:a}) into smaller ones
calculated over chirplet intervals i.e., $t_j \leq t_s k < t_{j+1}$ or
equivalently $jb \leq k \leq (j+1)b-1$ with $b=\delta_t/t_s$, the
number of samples in a chirplet interval. In the last sum, we separate
the terms corresponding to the transition between two consecutive
chirplets from the terms corresponding to the individual chirplets.

We now obtain a bound on each term of RHS of Eq.~(\ref{eq:c}),
starting with the first sum. Eq.~(\ref{eq:xi_zeta}) can be rewritten as
\be
\label{eq:zeta_cot}
\zeta_k=\frac{1}{2}[1+i\cot(d_k/2)].
\ee

The variables $\zeta_k$ are all located on the line $\Re(\zeta)=1/2$.

Within a chirplet interval, i.e. if $t_j \leq k t_s <t_{j+1}$, the
phase difference is a linear function of $k$ given by
\be
\label{eq:xi_k2}
d_k = 2 \pi t_s \left[(2-r) f_{m_j} + r f_{m_{j+1}}\right],
\ee
where $r=(2 t_{j,k} - t_s)/\delta_t$.  

We assume that the node frequencies of the CC are constrained in the
following bandwidth~:
\begin{equation}
f_l \leq f_{m_j} \leq f_s/2-f_l,
\end{equation}
where $f_l=f_s c/2$ and $0<c<1/2$. In other words, the CC cannot
approach arbitrarily close to neither DC nor the Nyquist frequency.

Since $0<r<2$, we have $4\pi t_s f_{m_j} \leq d_k \leq 4\pi t_s
f_{m_{j+1}}$ if $f_{m_{j}} \leq f_{m_{j+1}}$ (and the opposite in the
other case) which implies that
\begin{equation}
\label{constraint_d}
0 < 2\pi c \leq d_k \leq 2\pi (1-c) < 2\pi,
\end{equation}
for all $k$, hence $-\infty<\Im(\zeta_k)<+\infty$.

If $f_{m_{j}} \leq f_{m_{j+1}}$ (resp. $f_{m_{j}} \geq f_{m_{j+1}}$),
the phase difference $d_k$ and hence $\Im(\zeta_k)$, increases (resp.
decreases) monotonically with $k$.

Since their imaginary parts are finite and monotonic, the variables
$\zeta_k$ are associated to consecutive points on the line
$\Re(\zeta)=1/2$ of the complex plane. The sum of the lengths of the
segments linking two nearby points is equal to the length between the
extremes, thus
\begin{equation}
\sum_{k=jb+1}^{(j+1)b-1} |\zeta_{k+1} - \zeta_k| = |\zeta_{(j+1)b} - \zeta_{jb+1}|\,.
\end{equation}

Applying the mean value theorem to the function $g(x)=\cot(x/2)/2$,
whose derivative is $\dot{g}(x)=1/(4\sin^2(x/2))$ and using the
constraint in Eq.~ (\ref{constraint_d}), we obtain the following bound
\be
\label{eq:zeta2}
 |\zeta_{(j+1)b} - \zeta_{jb+1}| \leq \frac{|d_{(j+1)b} - d_{jb+1}|}{4\sin^2(\pi c)}.
\ee

We carry on by bounding the numerator
\begin{equation}
|d_{(j+1)b} - d_{jb+1}|=4 \pi t_s  |f_{m_{j+1}}-f_{m_j}| \leq 4\pi N_r'/N,
\end{equation}
and denominator with $2 c \leq \sin(\pi c)$ (this is valid for $0\leq
c \leq 1/2$) and by summing over all $j$ to finally obtain the bound
on first summation term in Eq.~(\ref{eq:c}),
\be
\label{eq:term2}
\sum_{j=0}^{N_t -1}~\sum_{k=jb}^{(j+1)b-1} |\zeta_{k+1} - \zeta_k| \leq \frac{\pi N_r' N_t}{4 N c^2}.
\ee

The second summation coming from the boundary points of the chirplet
intervals can be bounded in a similar way, considering that
\be
|\zeta_{jb+1} - \zeta_{jb}| \leq \frac{|d_{jb+1} - d_{jb}|}{4\sin^2(\pi c)},
\ee
and combining with
\be
|d_{jb+1}- d_{jb}| = 2\pi t_s^2 |f_{m_{j+1}}-f_{m_{j-1}}|/\delta_t \leq \frac{4 \pi t_s N_r'}{N \delta_t},
\ee
we get the result
\be
\label{eq:term3}
\sum_{j=1}^{N_t-2} |\zeta_{jb+1} - \zeta_{jb}| \leq \frac{\pi t_s N_r' N_t}{4 N c^2 \delta_t}.
\ee

Finally, from Eq.~(\ref{eq:zeta_cot}), we have the following inequalities
\begin{equation}
|\zeta_k|=\frac{1}{2|\sin(d_k/2)|}\leq \frac{1}{2\sin(\pi c)}\leq \frac{1}{4c},
\end{equation}
which, when applied with $k=1$ and $k=N-1$, set an upper limit to the
remaining terms in the RHS of Eq.~(\ref{eq:c}), noting that
$|1-\zeta_{N-1}|=|\zeta_{N-1}|$.

Combining this result with Eqs (\ref{eq:term2}) and (\ref{eq:term3}), we get
\be
|S| \leq \frac{1}{2 N c} + \frac{\pi N_r' N_t}{4 c^2 N^2} (1 + 1/b).
\ee

The number of samples in a chirplet interval being an integer $b\geq
1$, and selecting the dominating contribution, we conclude that $|S|
\lesssim \eta$ with
\be
\eta=\frac{\pi N_r' N_t}{2 c^2 N^2}.
\ee

This bound is obtained from a worst case estimate. Generally, $\delta$
and $\epsilon$ are smaller than this value. With the choice
of a small $c$, a more realistic estimate rather than a strict bound
can be obtained replacing the inequality $2c \leq \sin(\pi c)$ by the
first order Taylor approximation $\pi c \sim \sin(\pi c)$ in the proof
above, yielding the following estimate
\be 
\eta = \frac{N_r' N_t}{\pi c^2 N^2}.
\ee

Summarizing, we obtained an upper bound $\eta$ on $|S|$ by restricting
the frequency of the CC in a bandwidth defined by $c$. We rather use
the reciprocal i.e., we get the limits of the frequency bandwidth from
an acceptable value for $\eta$. If we assume that $N_r'\approx
4(N'/N_t)(N_f/(2N))$ as given by the regularity condition, the
frequency bandwidth is $[f_l,f_s/2-f_l]$ with
\begin{equation}
  f_l=\frac{f_s c}{2} \approx 2.5 \sqrt{N'}\delta_f \left(\frac{N_f}{N}\right)^{3/2} \left(\frac{0.1}{\eta}\right)^{1/2},
\end{equation}
where the leading constant is obtained from $\sqrt{20/\pi}\approx
2.5$.  We use this result in Sec. \ref{sec:approx1}.

\bibliography{paper}

\end{document}